\documentclass[%
 twocolumn,
 superscriptaddress,
 amsmath,amssymb,
 prx,
]{revtex4-2}

\usepackage{times}
\usepackage{color}
\usepackage{graphicx}
\usepackage[dvipsnames]{xcolor}
\usepackage{physics}
\usepackage{bm}
\usepackage{mathtools}
\usepackage{upgreek}
\usepackage[colorlinks=true,allcolors=blue]{hyperref}
\usepackage{footmisc}
\usepackage{makecell}
\usepackage{mathtools}
\usepackage{soul}

\newcommand{\kb}{k_\text{B}}

\begin{document}
\title{Positive- and negative-frequency noise from an ensemble of two-level fluctuators}

\author{Xinyuan You}
\affiliation{Northwestern--Fermilab Center for Applied Physics and Superconducting Technologies, Northwestern University, Evanston, Illinois 60208, USA}
\affiliation{Graduate Program in Applied Physics, Northwestern University, Evanston, Illinois 60208, USA}

\author{Aashish A. Clerk}
\affiliation{Pritzker School of Molecular Engineering, University of Chicago, 5640 South Ellis Avenue, Chicago, Illinois 60637, USA}

\author{Jens Koch}
\affiliation{Northwestern--Fermilab Center for Applied Physics and Superconducting Technologies, Northwestern University, Evanston, Illinois 60208, USA}
\affiliation{Department of Physics and Astronomy, Northwestern University, Evanston, Illinois 60208, USA}

\begin{abstract}
The analysis of charge noise based on the Bloch--Redfield treatment of an ensemble of dissipative two-level fluctuators generally results in a violation of the fluctuation--dissipation theorem. The standard Markov approximation (when applied to the two-level fluctuators coupled to a bath) can be identified as the main origin of this failure. The resulting decoherence rates only involve the bath response at the fluctuator frequency, and thus completely neglect the effects of frequency broadening. A systematic and computationally convenient way to overcome this issue is to employ the spectator-qubit method: by coupling an auxiliary qubit to the two-level fluctuator ensemble, an analytical approximation for $S(\omega)$ fully consistent with the fluctuation--dissipation theorem can be obtained. We discuss the resulting characteristics of the noise which exhibits distinct behavior over several frequency ranges, including a $1/f$ to $1/f^2$ crossover with a $T^3$ temperature dependence of the crossover frequency.
\end{abstract}

\maketitle

\section{introduction}
Random fluctuations of physical quantities in a qubit or its surrounding environment lead to decoherence limiting qubit performance. Common noise sources for superconducting qubits arise from fluctuating background charge~\cite{Nakamura2002,Astafiev2004}, magnetic flux~\cite{Yoshihara2006,Kumar2016b}, critical current \cite{VanHarlingen2004}, or quasiparticle poisoning~\cite{Martinis2009}. 
In widely used circuits such as the transmon~\cite{Koch2007} and fluxonium qubits~\cite{Manucharyan2009}, noise in different frequency ranges plays distinct roles in limiting coherence times: while dephasing rates are typically governed by low-frequency noise (e.g., $1/f$ noise), relaxation processes are usually dominated by  high-frequency noise (e.g., Nyquist noise).

This situation is altered in recent proposals for a new generation of qubits with intrinsic protection against noise, such as heavy fluxonium~\cite{Earnest2018,Nguyen2018}, the 0--$\pi$ qubit~\cite{Brooks2013,Groszkowski2017a,DiPaolo2018a,Gyenis2019}, and the current--mirror qubit~\cite{Kitaev2006,Weiss2019}. 
Qubits of this type are predicted to exhibit remarkably long coherence times due to the exponential suppression of transitions among the computational qubit states, achieved by localizing wavefunctions in separate regions of configuration space (disjoint support). 
Under these circumstances, depolarization is dominated by excitation processes producing leakage into higher qubit states beyond the computational subspace. The transition rates for such excitation processes, which involve energy transfer from the noise source to the qubit, are proportional to the noise spectral density $S(\omega)$, evaluated at \emph{negative frequencies}~\cite{aash_rmp}. Thus, the study of negative-frequency noise is crucial for understanding the depolarization of qubits with intrinsic protection. 

Here, we are particularly interested in the behavior of charge noise. While the microscopic origin of this noise has not been conclusively established~\cite{Clemens2019}, a number of theoretical studies have proceeded to consider an ensemble of two-level fluctuators (TLFs) as the cause of charge noise \cite{shnirman2005low,Schriefl2006,Constantin2009,Muller2015}. The predictions presented in these references for the positive-frequency noise  are consistent with a number of experimental observations \cite{Dutta1981,Weissman1988,Astafiev2004,Martinis2005,Astafiev2006,Paladino2014}. However, inspection of the noise spectral density derived from Bloch--Redfield theory reveals violations of the fluctuation--dissipation theorem \cite{PhysRev.83.34}. This theorem directly relates the negative-frequency part of $S(\omega)$ to its positive-frequency counterpart, or equivalently, the symmetrized spectral density to the imaginary part of a response function. To overcome this issue, we abandon Bloch--Redfield theory and instead extract $S(\omega)$ by computing the depolarization rate of an auxiliary qubit weakly coupled to the noise source. This \emph{spectator-qubit method} was  first introduced in the context of noise studies for single-electron transistors~\cite{Clerk2002,Schoelkopf2003}. The results thus derived for the charge-noise spectral density manifestly obey the fluctuation--dissipation theorem.

The paper is organized as follows. We describe the model of a single TLF weakly coupled to a thermal bath, and derive the corresponding spectral density in Sec.~\ref{sec:hamiltonian}. We then show in Sec.~\ref{sec:qrt} that results obtained from the Bloch--Redfield theory are inconsistent with the fluctuation--dissipation theorem. Our main results addressing this issue are presented in Secs.~\ref{sec:fdt} and \ref{sec:ens}, where we derive noise spectral densities first for a single TLF and then for an ensemble of TLFs. A crossover from $1/f$ to $1/f^2$ and further to Ohmic or white-noise behavior is predicted for positive frequencies, along with a corresponding exponentially suppressed negative-frequency component. We share our conclusions and outlook in Sec.~\ref{sec:con}, and provide additional details in the subsequent appendices.

\section{Two-level fluctuator coupled to a thermal bath}\label{sec:hamiltonian}
We start with a single TLF, and defer the case of an ensemble of TLFs to Sec.~\ref{sec:ens}. The Hamiltonian of a TLF coupled to a bosonic bath is
\begin{equation}\label{eq:ham}
    \hat{H}_\textrm{TLF--bath} = -\dfrac{1}{2} (\varepsilon\, \hat{\Sigma}_z + \Delta\, \hat{\Sigma}_x) + \hat{H}_\text{diss},
\end{equation}
where $\{\hat{\Sigma}_i\}$ is the set of Pauli operators associated with the TLF. Here, we adopt the notation commonly used for tunneling among the two lowest levels in an asymmetric double-well potential, with $\Delta$ the tunneling amplitude, and $\varepsilon$ the energy asymmetry. In principle, the TLF couples to the environment through both $\hat{\Sigma}_z$ and $\hat{\Sigma}_x$. However, the coupling via $\hat{\Sigma}_x$ is expected to be much smaller compared to the longitudinal coupling, and may be neglected~\cite{Temperature,Hunklinger1986,phillips1981amorphous}.
This leads to the Hamiltonian
\begin{equation}\label{eq:dissipation}
    \hat{H}_\text{diss} = \hat{\Sigma}_z\sum_\lambda (g_\lambda \hat{a}_\lambda + g_\lambda^* \hat{a}_\lambda^\dag) + \sum_\lambda \omega_\lambda \hat{a}^\dag_\lambda \hat{a}_\lambda
\end{equation}
describing the bath and its coupling to the TLF.
Here, the $\lambda$th mode of the bosonic bath has energy  $\omega_\lambda$, and couples to the TLF with coupling strength $g_\lambda$, through the ladder operators $\hat{a}_\lambda$, $\hat{a}_\lambda^\dag$. 
To characterize the effect of $\hat{H}_\text{diss}$, it is best to diagonalize the TLF Hamiltonian with the transformation
\begin{align}
    \hat{\Sigma}_z &= \cos(\theta) \,\hat{\sigma}_z - \sin(\theta)\, \hat{\sigma}_x, \\
    \hat{\Sigma}_x &= \sin(\theta) \,\hat{\sigma}_z + \cos(\theta) \, \hat{\sigma}_x.
\end{align}
The Pauli operators $\{ \hat{\sigma}_i \}$ refer to the eigenbasis of the TLF, and $\theta$ is defined by $\tan(\theta) = \Delta/\varepsilon$. The term proportional to $\hat{\sigma}_z$ generally introduces pure dephasing of the TLF. However, for the cubic bath spectral function considered throughout this paper, the pure-dephasing rate actually vanishes. Consequently, we may restrict our discussion to the case of purely transverse coupling~\footnote{In general, longitudinal coupling will modify the linewidth of $s_{xx}(\omega)$ in Eq.~\eqref{eq:spec}. The latter is responsible for the high-frequency regime of $S(\omega)$ calculated in Sec.~\ref{sec:ens}. Assuming that the TLFs are underdamped~\cite{shnirman2005low}, we have confirmed numerically that inclusion of longitudinal coupling does not lead to qualitative changes in the behavior of $S(\omega)$.}. Then, Eq.~\eqref{eq:ham} further simplifies to 
\begin{equation}\label{eq:diss}
    \hat{H}_\textrm{TLF--bath} = -\dfrac{1}{2}\omega_\text{t}\hat{\sigma}_z - \dfrac{\Delta}{\omega_\text{t}}\hat{\sigma}_x\sum_\lambda (g_\lambda \hat{a}_\lambda + g_\lambda^*\hat{a}_\lambda^\dag) + \sum_\lambda \omega_\lambda \hat{a}^\dag_\lambda \hat{a}_\lambda,
\end{equation} 
where $\omega_\text{t} = \sqrt{\varepsilon^2 + \Delta^2}$ is the eigenenergy of the TLF.

Within the master-equation formalism, used in later sections, the strength of dissipation is governed by the bath correlation function. In the Heisenberg picture, the bath operator coupling to the TLF is given by 
\begin{equation}
\hat{B}(t)=-\dfrac{\Delta}{\omega_\text{t}}\sum_\lambda (g_\lambda \hat{a}_\lambda e^{-i\omega_\lambda t}+ g_\lambda^*\hat{a}_\lambda^\dag e^{i\omega_\lambda t}).
\end{equation} 
In equilibrium, the correlation function is 
\begin{align}
    \begin{split}
      \langle \hat{B}(t) \hat{B}(0) \rangle = &\dfrac{\Delta^2}{\omega_\text{t}^2} \sum_\lambda |g_\lambda|^2 \big\{ [n_\text{B}(\omega_\lambda)+1] e^{-i\omega_\lambda t}  \\
      &+ n_\text{B}(\omega_\lambda) e^{i\omega_\lambda t} \big\},
    \end{split}{}
\end{align} 
where $n_\text{B}(\omega)=(e^{\beta\omega}-1)^{-1}$ is the Bose--Einstein distribution. Taking the Fourier transform of the correlation function, we obtain
\begin{align}
    \gamma(\omega) =& \,\int_{-\infty}^{+\infty} \mathrm{d}t \,e^{i\omega t} \langle \hat{B}(t) \hat{B}(0) \rangle \nonumber\\
    =& \, 2\pi \dfrac{\Delta^2}{\omega_\text{t}^2} \sum_\lambda |g_\lambda|^2 \big[(n_\text{B}(\omega_\lambda) +1)\delta(\omega-\omega_\lambda) \label{eq:bath_corr}\\
    &+ n_\text{B}(\omega_\lambda) \delta(\omega+\omega_\lambda)\big].\nonumber
\end{align}
Using the definition of the bath spectral function  $J(\omega)= \sum_\lambda |g_\lambda|^2 \delta(\omega-\omega_\lambda)$, the correlation function is 
\begin{equation}
    \gamma(\omega)=2\pi\dfrac{\Delta^2}{\omega_\text{t}^2}
    \begin{cases}
        J(\omega) (n_\text{B}(\omega) +1), & \omega \geq 0 \\
       J(-\omega) n_\text{B}(-\omega), & \omega < 0
    \end{cases}.
\end{equation}
The positive- and negative-frequency components of the correlation function can be interpreted as the golden-rule transition rates for the bath absorbing and emitting energy $|\omega|$, respectively. As expected, their magnitudes obey detailed balance 
\begin{equation}\label{eq:gamma_fdt}
    \gamma(-\omega)=\gamma(\omega) e^{-\beta\omega}.
\end{equation} 

Due to the coupling to the bath, a given TLF quantity $\hat{F}$ will undergo fluctuations. These fluctuations may be characterized by quoting the spectral density which is obtained as the Fourier transform of the autocorrelation function:
\begin{equation}\label{eq:s}
    s(\omega) = \int_{-\infty}^{+\infty} \mathrm{d}t \,e^{i\omega t} \big[\langle  \hat{F}(t) \hat{F}(0) \rangle - \langle  \hat{F} \rangle^2 \big].
\end{equation}
Here, $\hat{F}(t)$ denotes the Heisenberg representation, and $\langle \cdots \rangle$ refers to the quantum-mechanical expectation value in thermal equilibrium.
We are interested in the fluctuations of the dipole moment of the TLF, $p\hat{\Sigma}_\textrm{z}$~\cite{Faoro2006}, which we will relate to charge noise in Sec.\ \ref{sec:ens}.
Taking $\hat{F}$ to be $\hat{\Sigma}_z$, the spectral density in Eq.~\eqref{eq:s} can be rewritten as 
\begin{equation}\label{eq:spec}
    s(\omega) = \cos^2(\theta) s_{zz}(\omega) + \sin^2(\theta) s_{xx} (\omega),
\end{equation}
where   
\begin{equation}\label{eq:sa}
    s_{\alpha\alpha}(\omega ) = \int_{-\infty}^{+\infty} \mathrm{d}t \,e^{i\omega t} \big[ \langle  \hat{\sigma}_{\alpha}(t) \hat{\sigma}_\alpha(0) \rangle - \langle  \hat{\sigma}_\alpha \rangle^2 \big],
\end{equation}
with $\alpha = x,z$.
Note that the cross-correlations between $\hat{\sigma}_z$ and $\hat{\sigma}_x$ vanish for the TLF--bath coupling given in Eq.~\eqref{eq:diss}.

\section{Results from Bloch--Redfield theory}\label{sec:qrt}
We first follow Refs.~\onlinecite{shnirman2005low,Constantin2009} to calculate the spectral density of a TLF using Bloch--Redfield theory~\cite{bloch1957generalized,redfield1957theory}. The evolution of the expectation values of $\{\hat{\sigma}_i\}$ is governed by
\begin{align}
       \dfrac{\mathrm{d}}{\mathrm{d}t} \langle \hat{\sigma}_x(t) \rangle &= \phantom{-}\omega_\text{t}\langle \hat{\sigma}_y(t) \rangle - \gamma_2 \langle \hat{\sigma}_x(t) \rangle, \label{eq:qrt1}\\
        \dfrac{\mathrm{d}}{\mathrm{d}t} \langle \hat{\sigma}_y(t) \rangle &= -\omega_\text{t}\langle \hat{\sigma}_x(t) \rangle - \gamma_2 \langle \hat{\sigma}_y(t) \rangle, \label{eq:qrt2}\\ 
        \dfrac{\mathrm{d}}{\mathrm{d}t} \langle \hat{\sigma}_z(t) \rangle &=  - \gamma_1 \big(\langle \hat{\sigma}_z(t) \rangle -\langle \hat{\sigma}_z\rangle_\text{eq}\big),\label{eq:qrt3}
\end{align}
with $\gamma_1 = \gamma(\omega_\textrm{t}) + \gamma(-\omega_\textrm{t})$ denoting the depolarization rate, $\gamma_2 = \left[\gamma(\omega_\textrm{t}) + \gamma(-\omega_\textrm{t})\right]/2$ the dephasing rate, and $\langle \hat{\sigma}_z\rangle_\text{eq}= (\gamma_\uparrow - \gamma_\downarrow)/(\gamma_\uparrow + \gamma_\downarrow)$ the equilibrium polarization. For $t>0$, the solution to this system of differential equations is
\begin{align*}\nonumber
    \langle \hat{\sigma}_x(t)\rangle &= e^{-\gamma_2 t} \big[\langle \hat{\sigma}_y(0) \rangle \sin(\omega_\text{t} t) +\langle \hat{\sigma}_x(0) \rangle \cos(\omega_\text{t} t)\big],\\
    \langle \sigma_y(t)\rangle &= e^{-\gamma_2 t} \big[\langle \hat{\sigma}_y(0) \rangle \cos(\omega_\text{t} t)-\langle \hat{\sigma}_x(0) \rangle \sin(\omega_\text{t} t) \big],\\
    \langle \hat{\sigma}_z(t)\rangle &= e^{-\gamma_1 t} \big(\langle \hat{\sigma}_z(0) \rangle - \langle \hat{\sigma}_z\rangle_\text{eq} \big) +\langle \hat{\sigma}_z\rangle_\text{eq}.
\end{align*}
Employing the quantum regression theorem~\cite{breuer2002theory}, we further obtain the correlation functions 
\begin{align}
    \langle \hat{\sigma}_x(t)\hat{\sigma}_x(0) \rangle &= e^{-\gamma_2 t }\big[ \cos(\omega_\text{t} t) -i \langle \hat{\sigma}_z \rangle_\text{eq} \sin(\omega_\text{t} t)  \big], \label{eq:xx} \\ 
    \langle \hat{\sigma}_z(t)\hat{\sigma}_z(0) \rangle &= e^{-\gamma_1 t } (1 - \langle \hat{\sigma}_z \rangle_\text{eq}^2) +\langle \hat{\sigma}_z \rangle_\text{eq}^2. \label{eq:zz}
\end{align}
For the evaluation of the spectral density,  all expectation values and correlators above should be evaluated with respect to the equilibrium state. In this case, one finds $\langle \hat{\sigma}_x(0)\rangle = \langle \hat{\sigma}_y(0)\rangle =0$, and $\langle \hat{\sigma}_z(0)\rangle = \langle \hat{\sigma}_z \rangle_\text{eq}$. The negative-time counterparts to Eqs.~\eqref{eq:xx} and~\eqref{eq:zz}, required for the calculation of the spectral density, follow from $\langle \hat{A}_1(-t)\hat{A}_2(0) \rangle = \langle \hat{A}_1(t)\hat{A}_2(0) \rangle^*$~\cite{breuer2002theory}. 
The spectral density  of a single TLF~\cite{shnirman2005low,Constantin2009} can now be obtained via the correlation functions, and using Eq.~\eqref{eq:sa} along with a Fourier transform,
\begin{align}
    \begin{split}
        s^\text{BR}_{xx}(\omega) =& \,\dfrac{1+\langle \hat{\sigma}_z \rangle_\text{eq}}{2} \dfrac{2\gamma_2}{(\omega-\omega_\text{t})^2 + \gamma_2^2} \\ 
        &+ \dfrac{1-\langle \hat{\sigma}_z \rangle_\text{eq}}{2} \dfrac{2\gamma_2}{(\omega+\omega_\text{t})^2 + \gamma_2^2},\label{eq:sxx_c}
    \end{split}\\
    s^\text{BR}_{zz}(\omega) =& \,(1-\langle \hat{\sigma}_z \rangle_\text{eq} ^2) \dfrac{2\gamma_1}{\omega^2  + \gamma_1^2}\label{eq:szz_c}.
\end{align}
Here, the superscript BR refers to  Bloch--Redfield theory.

 The spectral density $s^\text{BR}_{zz}(\omega)$ is a Lorentzian centered at $\omega=0$ with linewidth $\gamma_1$. By contrast, $s^\text{BR}_{xx}(\omega)$ is a sum of two Lorentzians, centered at $\omega=\pm\omega_\text{t}$. The corresponding peak amplitudes are given by $(1\pm\langle \hat{\sigma}_z \rangle_\text{eq})/2$. Although, the  ratio of the peak heights can be confirmed to satisfy detailed balance,
\begin{equation}
    \dfrac{1+\langle \hat{\sigma}_z \rangle_\text{eq}}{1-\langle \hat{\sigma}_z \rangle_\text{eq}} = e^{\beta \omega_\text{t}},
\end{equation}
the overall profiles of $s_{xx}^\text{BR}(\omega)$ and $s_{zz}^\text{BR}(\omega)$ actually violate the fluctuation--dissipation theorem:
\begin{equation*}
    s^\text{BR}_{\alpha\alpha}(\omega)+ s^\text{BR}_{\alpha\alpha}(-\omega) \neq  \big[s^\text{BR}_{\alpha\alpha}(\omega)- s^\text{BR}_{\alpha\alpha}(-\omega)\big] \coth(\beta \omega /2),
\end{equation*}
with $\alpha=x,z$.
In particular, the right-hand side of the above equation vanishes for $s_{zz}^\textrm{BR}(\omega)$. Since this asymmetric part of the spectral density is directly related to the imaginary part of a Kubo response function~\cite{aash_rmp}, the Bloch--Redfield method fails to describe the response  of the TLF correctly. The origin of this failure can be understood as follows. 
As a result of the Markov approximation applied to the TLF--bath system, the rates of the bath-induced TLF depolarization and dephasing are determined by the bath correlation function exclusively evaluated at the system frequency $\pm\omega_\textrm{t}$. 
However, due to the TLF--bath interaction, the system frequency is actually broadened, which suggests that additional frequency components of $\gamma(\omega)$ in the vicinity of $\pm\omega_\textrm{t}$ may play a role. Including these frequency components turns out to be crucial in order to obtain a spectral density that obeys the fluctuation--dissipation theorem. 
The spectator-qubit method employed in the following sections applies the Markov approximation to an \emph{enlarged} system. This explicitly introduces the missing frequency components, and thus succeeds in restoring the fluctuation--dissipation theorem.

\section{Spectator-qubit method}\label{sec:fdt}
A more suitable method for obtaining the quantum noise spectral density consists of relating $S(\omega)$ to the dissipative dynamics of an auxiliary system weakly coupled to the noise source of interest. The simplest choice is a qubit acting as a noise spectrometer~\footnote{In principle, other systems such as a harmonic oscillator may be used as a probe. However, this would unnecessarily complicate the calculation of interest, and require additional considerations, e.g., of leakage into neighboring levels and varying matrix elements for different levels. Since the probe system here is merely a calculational tool, its particular nature is not of intrinsic interest and we thus choose the simplest possible system, i.e., a qubit.}. This description is of interest as an experimental  protocol, but is here employed exclusively as a convenient tool for computing the spectral density~\cite{Clerk2002,Schoelkopf2003}. Since the qubit fulfills the passive role of probing the noise source, we refer to this approach as the  spectator-qubit method. Within this approach, the TLF spectral density is derived from the depolarization rate of the spectator qubit. Applying a Markov approximation to the enlarged system of TLF and spectator qubit  induces contributions from a larger set of bath-correlator frequency components, which are no longer limited to the TLF frequency.
As discussed in Sec.~\ref{sec:qrt}, this enables us to steer clear of the issues plaguing the Bloch--Redfield theory, and derive results manifestly obeying the fluctuation--dissipation theorem.
While the presence of the spectator qubit is key to this method, we emphasize that the resulting noise spectral density is a property of the TLF only, and independent of the spectator qubit. The spectator qubit merely probes the noise spectral density of the TLF, and does not alter it.

\begin{figure}[tb]
    \includegraphics[width=0.48\textwidth]{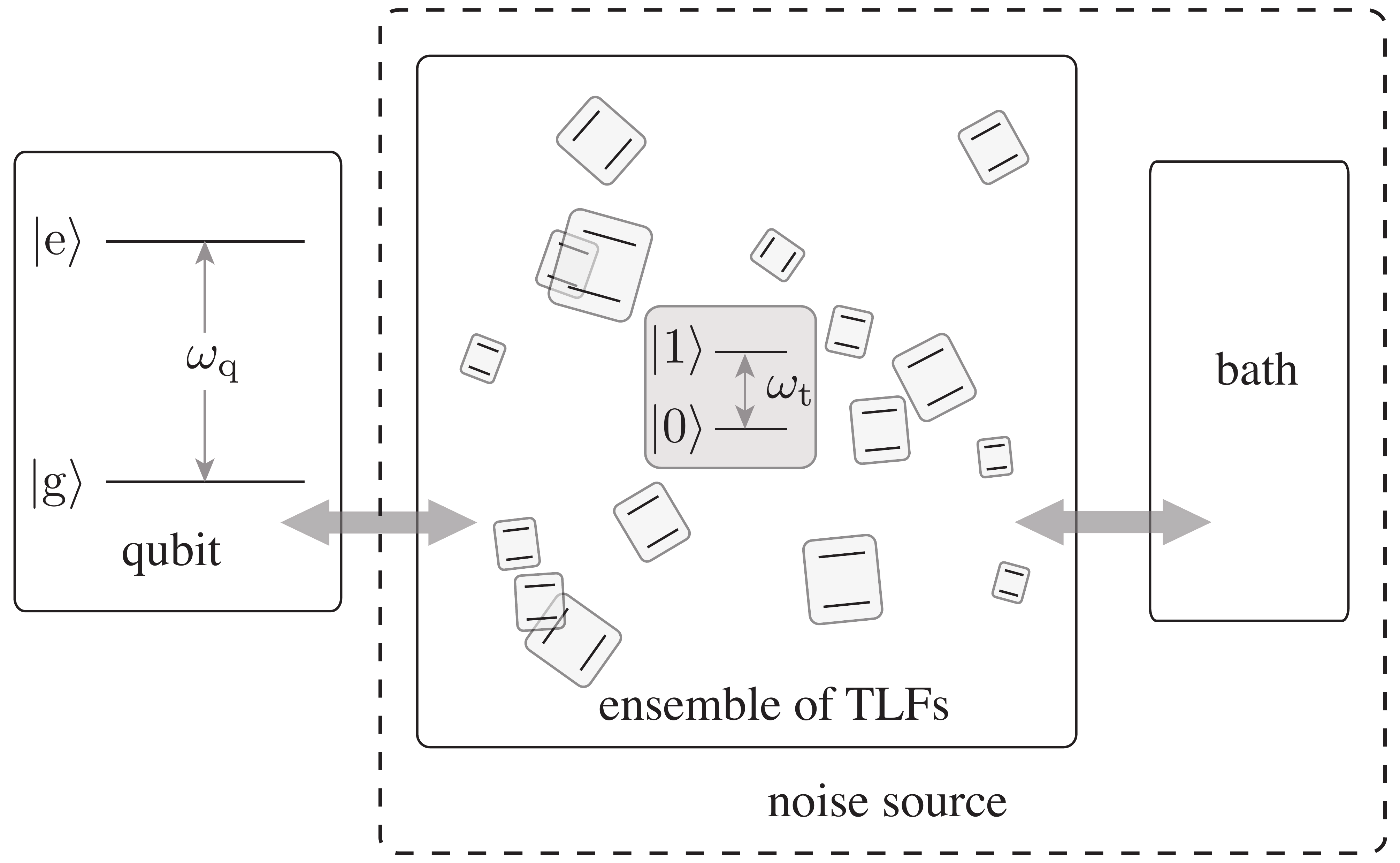}
    \caption{An ensemble of TLFs coupled to a bath. The ensemble and bath jointly act as a noise source that can be probed by an auxiliary qubit serving as a noise spectrometer.}
    \label{fig:sys}
\end{figure}
To implement the spectator-qubit method [Fig.~\ref{fig:sys}], we couple a TLF operator $\hat{\phi}(t)$ in the Heisenberg picture transversely to the qubit, as described by the Hamiltonian
\begin{equation}\label{eq:phi_dependence}
    \hat{H} = -\dfrac{1}{2} \omega_\text{q} \hat{\tau}_z + \kappa \hat{\tau}_x \hat{\phi}(t).
\end{equation}
Here, $\omega_\text{q}$ is the qubit energy and $\{ \hat{\tau}_i \}$ is the set of qubit Pauli operators. The coupling between the qubit and TLF is parametrized by $\kappa$. The spectral density $s_{\phi\phi}(\omega)$ of the noise can be extracted from the relaxation and excitation rates of the qubit. For $\kappa/\omega_\text{q}\ll1$, Fermi's golden rule yields
\begin{equation}\label{eq:gold}
     s_{\phi\phi}(+\omega_\text{q}) = \kappa^{-2}\Gamma_{\downarrow}, \qquad
     s_{\phi\phi}(-\omega_\text{q}) = \kappa^{-2}\Gamma_{\uparrow},
\end{equation} 
where $\Gamma_\downarrow$ and $\Gamma_\uparrow$ are the qubit relaxation and excitation rates. In particular, we are interested in the spectral densities $s_{zz}(\omega)$ and $s_{xx}(\omega)$, for $\hat{\phi}=\hat{\sigma}_z,\hat{\sigma}_x$ [see Eq.~\eqref{eq:spec}]. In the following, we first calculate the qubit depolarization rate, and then analyze the noise spectral density resulting from Eq.~\eqref{eq:gold}.

\subsection{Depolarization rate of a qubit coupled with a TLF}\label{subsec:qubitandtlf}
Depolarization of the qubit arises from coupling to the TLF. The Hamiltonian for qubit, TLF and bath is $\hat{H}_\text{S}+\hat{H}_\text{I}+\hat{H}_\text{B}$,
where $\hat{H}_\text{S} = - \dfrac{1}{2}\omega_\text{q}\hat{\tau}_z + \kappa\hat{\tau}_x \hat{\phi} - \dfrac{1}{2}\omega_\text{t}\hat{\sigma}_z$ describes the combined system of qubit and TLF, $\hat{H}_\text{B} = \sum_\lambda \omega_\lambda \hat{a}^\dag_\lambda \hat{a}_\lambda$ captures the bath modes, and $\hat{H}_\text{I} =-\frac{\Delta}{\omega_\text{t}}\hat{\sigma}_x\sum_\lambda (g_\lambda \hat{a}_\lambda + g_\lambda^*\hat{a}_\lambda^\dag)$ denotes the TLF--bath interaction. To model the depolarization dynamics of the qubit, we derive a suitable master equation.
While we largely follow the standard derivation~\cite{breuer2002theory}, there are several crucial differences discussed in the following~\footnote{The same approach applied to the problem of a single-electron transistor~\cite{Schoelkopf2003} generates a spectral density that is identical to the result obtained by a much more intricate calculation based on Keldysh diagrammatics~\cite{Johansson2002}. }.

\paragraph{Markov approximation in the Schr\"{o}dinger picture.} 
The Markov approximation is commonly applied to convert a time-nonlocal equation to a time-local one, where the density matrix at retarded time is replaced with that at the present time~\cite{breuer2002theory}. This replacement is appropriate if the system dynamics is slow compared to the bath correlation time. It is important to note that the dynamical time scales present in the dynamics of the density matrix crucially depend on whether we employ the Schr\"{o}dinger or interaction picture. Here the dynamics of interest is governed by the depolarization of the qubit. While the Schr\"{o}dinger picture directly reveals this process, the interaction picture leads to a combination of depolarization and fast oscillatory behavior (see Appendix~\ref{app:markov}). Therefore, in this specific case, we perform the Markov approximation in the Schr\"{o}dinger picture instead of the interaction picture, as is the most common choice~\cite{breuer2002theory}.

\paragraph{Omitting the secular approximation.}
Usually, the secular approximation is invoked next in order to cast the master equation into Lindblad form~\cite{breuer2002theory}. This strategy does not succeed here: application of the  Markov approximation in the Schr\"odinger picture (rather than the interaction picture) invariably leads to a non-Lindblad master equation. As a result, there is no advantage in applying the secular approximation, and we hence choose not to apply it and retain all contributions. 

With the Markov approximation in the Schr\"{o}dinger picture and in the absence of the secular approximation, the master equation takes the form of
\begin{align}
    \dfrac{\mathrm{d}}{\mathrm{d}t} \hat{\rho}(t) =& \,-i [\hat{H}_\text{S}, \hat{\rho}(t)] \nonumber\\
    & + \sum_{ij} \dfrac{1}{2} \gamma(-\varepsilon_{ij}) \big(\hat{\Pi}_i \hat{\sigma}_x \hat{\rho}(t) \hat{\Pi}_j \hat{\sigma}_x - \hat{\sigma}_x \hat{\Pi}_i \hat{\sigma}_x \hat{\rho}(t) \hat{\Pi}_j \nonumber\\
    &  + \hat{\sigma}_x \hat{\Pi}_j \hat{\rho}(t) \hat{\sigma}_x \hat{\Pi}_i - \hat{\Pi}_j\hat{\rho} (t) \hat{\sigma}_x \hat{\Pi}_i \hat{\sigma}_x \big).\label{eq:master}
\end{align}
Here, $\hat{\rho}(t)$ is the system density matrix, $\hat{\Pi}_i$ is a projector onto the $i$th eigenstate of $\hat{H}_\text{S}$, $\varepsilon_{ij}$ is the energy difference between the $i$th and $j$th eigenstates, and $\gamma(\omega)$ is the Fourier transform of the bath correlation function defined in Eq.~\eqref{eq:bath_corr}. 
Comparison of the master equation \eqref{eq:master} with the Lindblad master equation from Bloch--Redfield theory shows that the spectator-qubit method indeed produces damping terms that involve additional frequency components of the bath correlation function, specifically  $\gamma(\pm\omega_\textrm{t}\pm\omega_\textrm{q})$.
To keep notation compact, we introduce the following abbreviations for the rates:
\begin{align}\label{eq:gamma_def}
    \begin{split}
        & \gamma(\omega_\text{t}) = \gamma_\downarrow,\qquad\qquad\,\,\, \qquad \gamma(-\omega_\text{t}) = \gamma_\uparrow, \\
        & \gamma(\omega_\text{q}) = \gamma^+,\qquad\qquad\,\qquad  \gamma(-\omega_\text{q}) = \gamma^-, \\
        & \gamma(\omega_\text{t}-\omega_\text{q}) = \gamma_\downarrow^-, \qquad\qquad \gamma(-\omega_\text{t}+\omega_\text{q}) = \gamma_\uparrow^+, \\ 
        &\gamma(\omega_\text{t}+\omega_\text{q}) = \gamma_\downarrow^+, \qquad\qquad \gamma(-\omega_\text{t}-\omega_\text{q}) = \gamma_\uparrow^-.
    \end{split}
\end{align}

Since we are only interested in the limit of weak coupling between spectator qubit and TLF (i.e., $\kappa\to 0$), we only need to solve for the depolarization dynamics of the qubit perturbatively.
We denote the quantum numbers of the qubit by $\{\text{g},\text{e}\}$, and those of the TLF by $\{0,1\}$. It is convenient to convert the reduced density matrix of the combined system
\begin{equation}
   \hat{\rho}= \left(
\begin{array}{cccc}
 \text{$\rho_{\text{ee}11} $} & \text{$\rho_{\text{ee}10} $} & \text{$\rho_{\text{eg}11} $} & \text{$\rho_{\text{eg}10} $} \\
 \text{$\rho_{\text{ee}01} $} & \text{$\rho_{\text{ee}00} $} & \text{$\rho_{\text{eg}01} $} & \text{$\rho_{\text{eg}00} $} \\
 \text{$\rho_{\text{ge}11} $} & \text{$\rho_{\text{ge}10} $} & \text{$\rho_{\text{gg}11} $} & \text{$\rho_{\text{gg}10} $} \\
 \text{$\rho_{\text{ge}01} $} & \text{$\rho_{\text{ge}00} $} & \text{$\rho_{\text{gg}01} $} & \text{$\rho_{\text{gg}00} $} \\
\end{array}
\right),
\end{equation}
into coherence vector form $| \rho )$. In this way, the master equation can be written as
\begin{equation}\label{eq:evo_mat}
    \dfrac{\mathrm{d}}{\mathrm{d}t} | \rho )= \Lambda | \rho ),
\end{equation}
where $\Lambda$ is a $16\times16$ matrix. Treating the coupling between qubit and TLF perturbatively, we expand the evolution matrix in powers of $\kappa$:
\begin{equation}\label{eq:lambda_evo}
    \Lambda = \Lambda_0 + \kappa \Lambda_1 + \kappa^2 \Lambda_2 + \mathcal{O}(\kappa^3).
\end{equation} 
The dynamics of the uncoupled system is determined by $\Lambda_0$, the explicit form of which is given in Appendix~\ref{app:evo_mat}. The stationary states of the system are given by the two zero modes of $\Lambda_0$
\begin{align}
   \hat{\rho}_\text{g}= \left(
\begin{array}{cc}
 0 & 0 \\
 0 & 1
\end{array}
\right) 
\otimes 
\left(
\begin{array}{cc}
 p_1^\text{eq} & 0 \\
 0 & p_0^\text{eq}
\end{array}
\right) ,
\label{eq:zero_1}\\
\hat{\rho}_\text{e}= \left(
\begin{array}{cc}
 1 & 0 \\
 0 & 0
\end{array}
\right) 
\otimes 
\left(
\begin{array}{cc}
 p_1^\text{eq} & 0 \\
 0 & p_0^\text{eq}
\end{array}
\right),
\label{eq:zero_2}
\end{align}
which place the qubit in the ground or excited state. The TLF occupies the equilibrium state, characterized by the probabilities
\begin{equation}
    p_0^\text{eq} = \dfrac{\gamma_\downarrow}{\gamma_\uparrow+\gamma_\downarrow},
    \quad 
    p_1^\text{eq} = \dfrac{\gamma_\uparrow}{\gamma_\uparrow+\gamma_\downarrow}.
\end{equation} 

Second-order perturbation theory in the qubit--TLF coupling induces transitions between the two zero modes. To facilitate the degenerate-perturbative calculation, we define the projector onto the degenerate subspace~\cite{Brody2014}:
\begin{equation}\label{eq:proj}
    P=\dfrac{|\rho_\text{g}) (\phi_\text{g}|}{(\phi_\text{g}|\rho_\text{g})} + \dfrac{|\rho_\text{e}) (\phi_\text{e}|}{(\phi_\text{e}|\rho_\text{e})}.
\end{equation}
Here $(\phi_\text{g}|$ and $(\phi_\text{e}|$ are the two left eigenvectors of $\Lambda_0$ with eigenvalue zero.
Since $\Lambda_1$ vanishes in the degenerate subspace, $P\Lambda_1P=0$, there is no first-order correction. The leading corrections are hence of second order, and arise from both $\Lambda_1$ and  $\Lambda_2$. As shown in Appendix~\ref{app:dept}, the perturbative treatment leads to an effective evolution matrix:
\begin{equation}\label{eq:pt}
    \Lambda_\textrm{m} = P\Lambda_2 P
    -P\Lambda_1 (\mathrm{1}-P)\Lambda_0^{-1} (\mathrm{1}-P) \Lambda_1 P.
\end{equation}
Both terms in the above equation act on states in the degenerate subspace. While the first term simply applies $\Lambda_2$, the second term involves an intermediate projection $\mathrm{1}-P$ onto the complementary subspace, after applying $\Lambda_1$. The operator $\Lambda_0^{-1}$ plays an analogous role as $(\hat{H}-E)^{-1}$ which generates the energy denominators in ordinary perturbation theory.
Employing $\Lambda_\text{m}$, the relaxation and excitation rates can be obtained via 
\begin{equation}\label{eq:depolarization}
    \Gamma_\downarrow = \kappa^2( \phi_\text{g} | \Lambda_\textrm{m} | \rho_\text{e}), \qquad \Gamma_\uparrow = \kappa^2( \phi_\text{e} | \Lambda_\textrm{m} | \rho_\text{g}). 
\end{equation}
Note that the matrix $\Lambda_\textrm{m}$ depends on the TLF operator $\hat{\phi}$ that couples to the qubit [see Eq.~\eqref{eq:phi_dependence}]. For the case $\hat{\phi}=\hat{\sigma}_z$, the transition rates are 
\begin{align}
    \Gamma_\uparrow^z &= \kappa^2 \dfrac{4(p_1^\text{eq}\gamma_\downarrow^-+p_0^\text{eq}\gamma_\uparrow^-)}{\omega_\text{q}^2 + (\gamma_\downarrow^++\gamma_\downarrow^- + \gamma_\uparrow^++\gamma_\uparrow^-)^2/4},\label{eq:szup}\\ 
    \Gamma_\downarrow^z &= \kappa^2 \dfrac{4(p_1^\text{eq}\gamma_\downarrow^++p_0^\text{eq}\gamma_\uparrow^+)}{\omega_\text{q}^2 + (\gamma_\downarrow^++\gamma_\downarrow^- + \gamma_\uparrow^++\gamma_\uparrow^-)^2/4}.\label{eq:szdown}
\end{align}
For the case $\hat{\phi}=\hat{\sigma}_x$, the transition rates are then
\begin{align}
    \Gamma_\uparrow^x &= \kappa^2\dfrac{4\omega_\text{t}^2 \gamma^- }{(\omega_\text{q}^2-\omega_\text{t}^2)^2 + \omega_\text{q}^2 (\gamma^+ +\gamma^-)^2},\label{eq:sxup}\\ 
    \Gamma_\downarrow^x &= \kappa^2\dfrac{4\omega_\text{t}^2 \gamma^+  }{(\omega_\text{q}^2-\omega_\text{t}^2)^2 + \omega_\text{q}^2 (\gamma^+ +\gamma^-)^2}.\label{eq:sxdown}
\end{align}

\subsection{Noise spectral density of a single TLF}
To obtain the noise spectral density of a single TLF, we need to calculate $s_{zz}(\omega)$ and $s_{xx}(\omega)$, respectively [see Eq.~\eqref{eq:spec}]. These spectral densities can be calculated from the depolarization rates derived above, using Eq.~\eqref{eq:gold}.

\subsubsection{Noise spectral density $s_{zz}(\omega)$ of a TLF}\label{sec:szz}
\begin{figure}
    \includegraphics[width=0.5\textwidth]{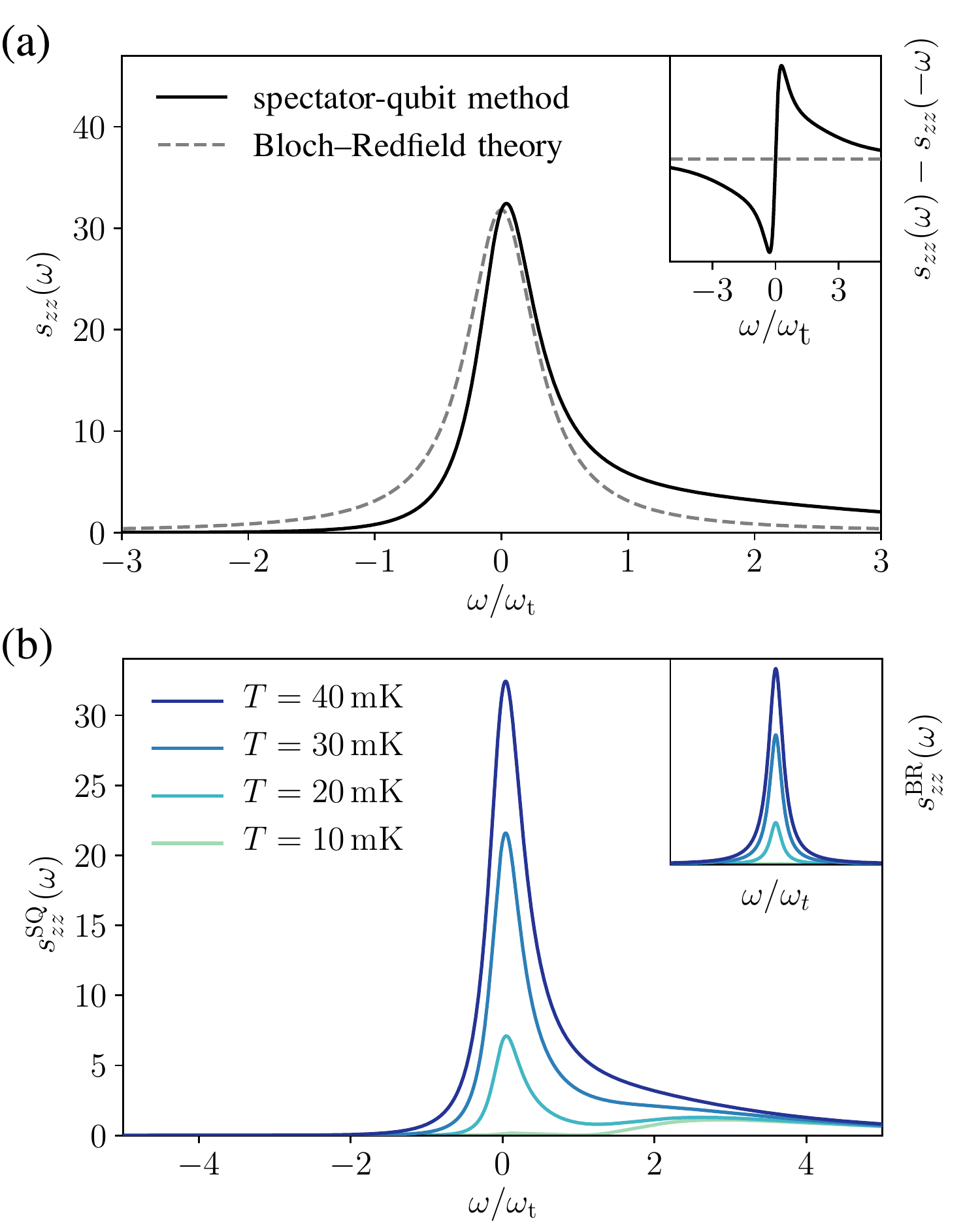}
    \caption{
    (a) Comparison between noise spectral densities $s_{zz}(\omega)$ for a single TLF, calculated in two different ways. The solid black curve represents the results from the qubit-as-spectrometer approach. For comparison, the dashed curve shows $s_{zz}(\omega)$ computed via the Bloch--Redfield theory. The inset shows the asymmetric parts of the spectral densities. 
    (b) Noise spectral density $s_{zz}^\text{SQ}(\omega)$ for a single TLF, evaluated at different temperatures. While the low-frequency part is strongly suppressed, the high-frequency part is insensitive to the lowering of the temperature. This behavior can be explained by an analysis of the underlying perturbative processes. The inset shows temperature dependence of $s_{zz}^\text{BR}(\omega)$ obtained from the Bloch--Redfield theory.
    [Parameters used: $T=40\, \text{mK}$ in (a), $k_\text{B}^2 J_0 \Delta^2 / \omega_\text{t}^2 =6.25\,\text{K}^{-2}$, $\omega_\text{D}/k_\text{B}=470\,\textrm{K}$, and $\omega_\text{t}/k_\text{B}=0.08\,\textrm{K}$.] }
    \label{fig:szz}
\end{figure}
% \begin{figure}    
%     \includegraphics[width=0.5\textwidth]{fig_szz_temp.pdf}
%     \caption{Noise spectral density $s_{zz}^\text{SQ}(\omega)$ for a single TLF, evaluated at different temperatures. While the low-frequency part is strongly suppressed, the high-frequency part is insensitive to the lowering of the temperature. This behavior can be explained by an analysis of the underlying perturbative processes. The inset shows temperature dependence of $s_{zz}^\text{BR}(\omega)$ obtained from the Bloch--Redfield theory. [Parameters used: $k_\text{B}^2 J_0 \Delta^2 / \omega_\text{t}^2 =6.25\,\text{K}^{-2}$, $\omega_\text{D}/k_\text{B}=470\,\textrm{K}$, and $\omega_\text{t}/k_\text{B}=0.08\,\textrm{K}$.]}
%     \label{fig:szz_temp}
% \end{figure}
\begin{figure*}
    \includegraphics[width=0.9\textwidth]{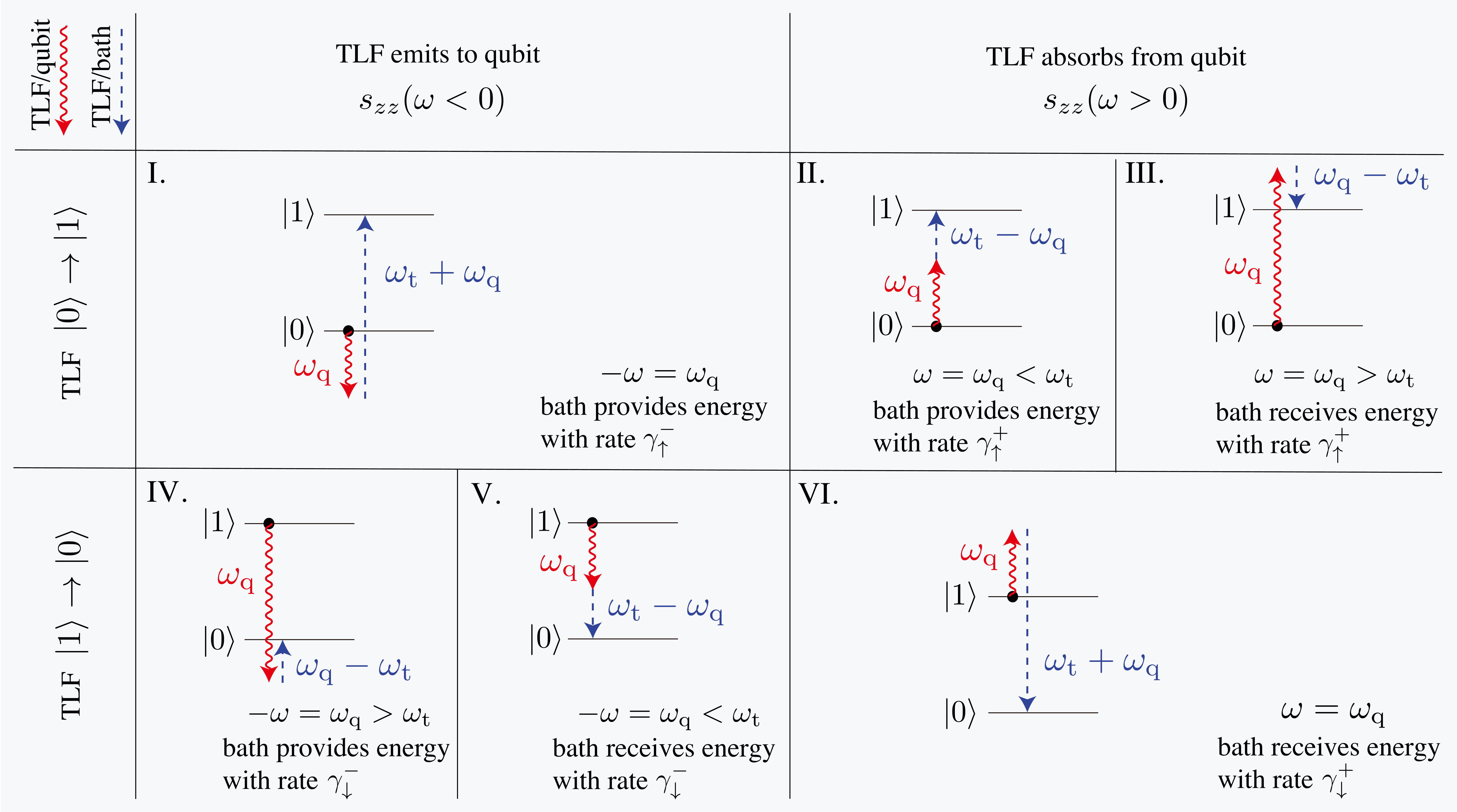}
    \caption{Relevant leading-order processes contributing to the positive-frequency (right column) and negative-frequency (left column) noise spectral density. Each process involves the transition of the TLF ($|0\rangle\to|1\rangle$ or $|1\rangle\to|0\rangle$), which is correlated with qubit absorption (downward red wiggly arrow) or emission (upward red wiggly arrow). Note that the qubit frequency here resembles the external frequency entering the bubble diagram, when calculated diagrammatically~\cite{Johansson2002}. The bath provides (upward blue dashed arrow) or receives (downward blue dashed arrow) the energy mismatch between the qubit and TLF.}
    \label{fig:process}
\end{figure*}
First, we calculate $s_{zz}(\omega)$. Employing Eq.~\eqref{eq:gold}, the values of the spectral density at $\pm\omega_\text{q}$ are obtained from depolarization rates Eqs.~\eqref{eq:szup} and \eqref{eq:szdown}
\begin{align}
    s_{zz}(+\omega_\text{q}) &= \dfrac{4(p_1^\text{eq}\gamma_\downarrow^++p_0^\text{eq}\gamma_\uparrow^+)}{\omega_\text{q}^2 + (\gamma_\downarrow^++\gamma_\downarrow^- + \gamma_\uparrow^++\gamma_\uparrow^-)^2/4}, \label{eq:szp}\\
    s_{zz}(-\omega_\text{q}) &= \dfrac{4(p_1^\text{eq}\gamma_\downarrow^-+p_0^\text{eq}\gamma_\uparrow^-)}{\omega_\text{q}^2 + (\gamma_\downarrow^++\gamma_\downarrow^- + \gamma_\uparrow^++\gamma_\uparrow^-)^2/4}.\label{eq:szn}
\end{align}
By treating the qubit frequency $\omega_\text{q}$ as a sweeping parameter, all positive- and negative-frequency components of the noise spectral density are obtained. Eqs.~\eqref{eq:szp} and \eqref{eq:szn} can be further combined into one compact expression valid for both positive and negative $\omega$~\footnote{The form of the obtained spectral density in Eq.~\eqref{eq:szz} is analogous to $s_{zz}(\omega)$ for a single-electron transistor~\cite{Johansson2002,Schoelkopf2003}. While the bath interacting with the single-electron transistor is fermionic, the relevant excitations are electron--hole pairs characterized by a Bose--Einstein distribution. It is thus plausible that the two cases lead to similar expressions.}:
\begin{equation}\label{eq:szz}
    s_{zz}(\omega) = \dfrac{4(p_1^\text{eq}\gamma_\downarrow^++p_0^\text{eq}\gamma_\uparrow^+)}{\omega^2 + (\gamma_\downarrow^++\gamma_\downarrow^- + \gamma_\uparrow^++\gamma_\uparrow^-)^2/4}.
\end{equation}
Recall that the rates $\gamma_{\uparrow\downarrow}^\pm$ depend on $\omega$ [see Eq.~\eqref{eq:gamma_def}] and obey detailed balance [Eq.~\eqref{eq:gamma_fdt}]. As a result, we find that $s_{zz}(-\omega)= s_{zz}(\omega)e^{-\beta\omega}$, so there is no violation of the fluctuation--dissipation theorem. To distinguish this spectral density from $s^\text{BR}_{zz}(\omega)$ obtained in Sec.~\ref{sec:qrt}, we refer to Eq.~\eqref{eq:szz} as $s^\text{SQ}_{zz}(\omega)$ in the following, where the superscript SQ stands for spectator-qubit method.

To evaluate and compare $s^\text{SQ}_{zz}(\omega)$ and $s^\text{BR}_{zz}(\omega)$, the bath spectral function $J(\omega)$ entering the rates $\gamma_{\uparrow\downarrow}^\pm$ must be specified. Here, we consider a cubic spectral function with exponential cutoff~\footnote{For a cubic bath spectral function, the TLF's pure-dephasing rate vanishes, $\gamma(0)=0$.},
\begin{equation}\label{eq:jw}
    J(\omega) = J_0 \,\omega^3 \,e^{-\omega^2/2\omega_\text{D}^2},
\end{equation}
where $J_0$ characterizes the interaction strength between the bath and the system, and $\omega_\text{D}$ denotes the high-frequency cutoff.
Figure~\ref{fig:szz}(a) shows an example comparing the resulting two spectral densities. While $s_{zz}^\text{BR}(\omega)$ has the shape of a symmetric Lorentzian, $s_{zz}^\text{SQ}(\omega)$ exhibits an asymmetric profile, consistent with the requirement from the fluctuation--dissipation theorem. Figure~\ref{fig:szz}(a) also plots the asymmetric-in-frequency part of the quantum noise spectral density.  Using standard linear response theory, this is equal to the negative imaginary part of the Kubo susceptibility $\chi_{zz}[\omega]$ (up to a factor $1/2 $), where
\begin{equation}
    \chi_{zz}(t) \equiv -i \theta(t) \langle 
    \left[ \hat{\sigma}_z(t), \hat{\sigma}_z(0) \right] \rangle.
\end{equation}
This susceptibility describes the linear response of $\langle \hat{\sigma}_z(t) \rangle$ to a time-varying perturbation that couples to $\hat{\sigma}_z$.  Thus, the quantum part of our spectral density is directly related to the out-of-phase response of our system to a time-varying perturbation.  This dissipative response is maximal when $\omega$ approximately matches the total TLF relaxation rate, a phenomenon that is well known in other contexts (e.g., in the Zener model of anelasticity \cite{Zener1948}).

More significant deviations between $s^\text{BR}_{zz}(\omega)$ and $s^\text{SQ}_{zz}(\omega)$ emerge when inspecting the behavior of the noise spectral density at different temperatures [see Fig.~\ref{fig:szz}(b)]. While $s^\text{BR}_{zz}(\omega)$ is uniformly suppressed when lowering the temperature, $s^\text{SQ}_{zz}(\omega)$ shows a richer behavior as a function of temperature. Specifically, we observe that all curves rapidly converge to a single asymptote $\propto 1/\omega^3$ in the high-frequency limit $\omega \gg \omega_\text{t}$, indicating a strong suppression of temperature dependence in this regime. In the low-frequency region of the spectrum ($\omega < \omega_\text{t}$), decreasing the temperature strongly suppresses the central peak, intermediately producing a double-peaked shape before reducing again to a single peak. In order to elucidate this temperature dependence, we first introduce the perturbative processes induced by the TLF--qubit coupling~\footnote{Eq.~\eqref{eq:master} is not in the Lindblad form, which in principle disables the unravelling of the master equation with quantum trajectory theory. However, the notion of processes can still be established from the more complicated diagrammatic approach, see Ref.~\onlinecite{Johansson2002} for example.} and then analyze their frequency and temperature dependence.

The relevant set of perturbative processes depends crucially on the nature of the coupling $\kappa \hat{A}_\text{qubit} \hat{B}_\text{TLF}$. In the framework of employing the spectator-qubit method, the qubit coupling operator is fixed to $\hat{A}_\text{qubit}=\hat{\tau}_x$, while the TLF operator is chosen according to the noise spectral density of interest, namely $\hat{B}_\text{TLF}=\hat{\sigma}_z$ for $s_{zz}(\omega)$ and $\hat{B}_\text{TLF}=\hat{\sigma}_x$ for $s_{xx}(\omega)$. We first focus on $s_{zz}(\omega)$, in which case the perturbative treatment of the coupling results in four different processes involving excitation and relaxation of TLF and/or qubit. The energy mismatch between qubit and TLF is compensated by additional energy emission or absorption due to the bath, leading to a total of six processes shown in Fig.~\ref{fig:process}~\footnote{Two of the eight possible combinations are ruled out by energy conservation.}. The positive-frequency behavior of the noise spectral density [Fig.~\ref{fig:szz}(b)] is determined by the processes II, III and VI (i.e., right column of Fig.~\ref{fig:process}) which involve energy transfer from the qubit to the TLF. In the following, we study the frequency and temperature dependence of those processes. 

\paragraph{Frequency-dependent switching between processes.} Not all three processes are active for all positive frequencies. 
While process VI leading to TLF relaxation occurs for all positive frequencies, processes II and III causing TLF excitation are mutually exclusive. Process II takes place whenever $\omega=\omega_\text{q}<\omega_\text{t}$, in which case qubit relaxation does not provide sufficient energy for exciting the TLF (red wiggly line in II, Fig.~\ref{fig:process}), and the bath has to provide the required additional energy $\omega_\text{t}-\omega_\text{q}$ (blue dashed line in II). By contrast, process III occurs in the opposite situation $\omega=\omega_\text{q}>\omega_\text{t}$ when qubit relaxation leads to an energy excess that is absorbed by the bath. (Similar threshold behavior is found in the context of noise from a single-electron transistor~\cite{Johansson2002,Schoelkopf2003}.)

\paragraph{Temperature dependence.} The temperature dependence of the rates for these processes differs characteristically. Processes II and VI are both suppressed as temperature is lowered: II requires thermal emission from the bath and is accompanied by a thermal factor of $n_\text{B}(\omega_\text{t}-\omega_\text{q})$; VI necessitates initial population of the TLF which is associated with a Boltzmann factor of $e^{-\omega_\text{t}/\kb T}$. On the other hand, process III is only weakly dependent on temperature, since it requires neither emission from the bath nor thermal excitation of the TLF. 

Considering both aspects of frequency-dependent switching and temperature dependence of processes and their associated rates, one concludes the following. For low frequencies $\omega < \omega_\text{t}$, the two active processes II and VI both undergo strong suppression with decreasing temperature, explaining the suppression of the central peak in $s_{zz}(\omega<\omega_\text{t})$ as temperature is lowered. At higher frequencies $\omega > \omega_\text{t}$ and temperatures $\kb T < \omega_\text{t}$, process III dominates over VI. Together with the weak temperature dependence of process III, this explains the convergence of $s_{zz}(\omega>\omega_\text{t})$ at different temperatures to a common asymptote.

\subsubsection{Noise spectral density $s_{xx}(\omega)$ of a TLF}
\begin{figure}
    \includegraphics[width=0.5\textwidth]{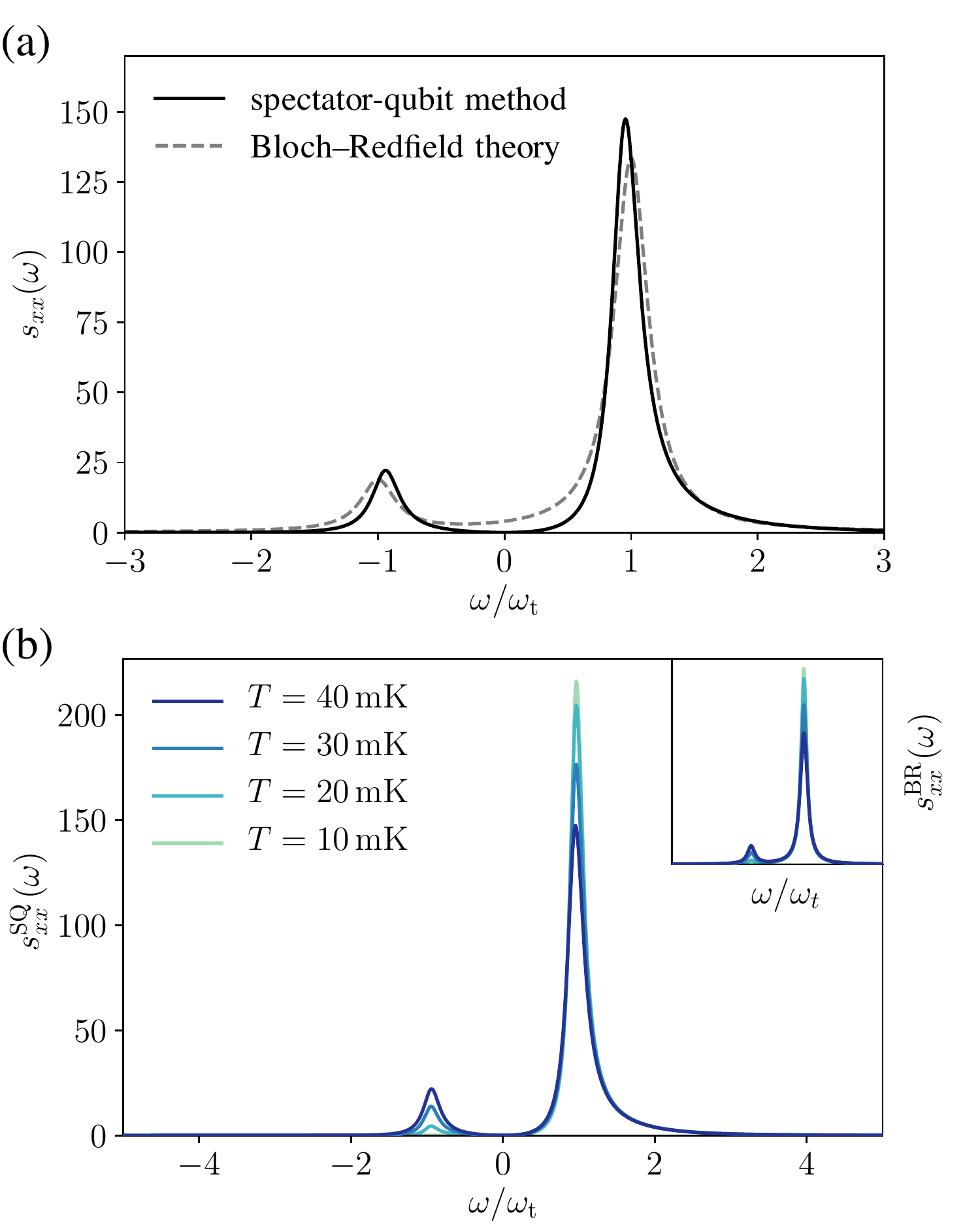}
    \caption{
    (a) Comparison between noise spectral densities $s_{xx}(\omega)$ for a single TLF, calculated in two different ways. The solid black curve represents the results from the qubit-as-spectrometer approach. For comparison, the dashed curve shows $s_{xx}(\omega)$ computed via the Bloch--Redfield theory. 
    (b) Noise spectral density $s_{xx}^\text{SQ}(\omega)$ for a single TLF, evaluated at different temperatures. The negative peak is exponentially suppressed as temperature is lowered. Away from the peak, the positive-frequency spectral density is insensitive to the lowering of temperature. When close to the peak, $s_{xx}^\text{SQ}(\omega)$ is enhanced due to the Zeno effect. The inset shows temperature dependence of $s_{xx}^\text{BR}(\omega)$ obtained from the Bloch--Redfield theory.
    [Parameters used: $T=40\, \text{mK}$ in (a), $k_\text{B}^2 J_0 \Delta^2 / \omega_\text{t}^2 =6.25\,\text{K}^{-2}$, $\omega_\text{D}/k_\text{B}=470\,\textrm{K}$, $\omega_\text{t}/k_\text{B}=0.08\,\textrm{K}$.]}
    \label{fig:sxx}
\end{figure}
% \begin{figure}
%     \includegraphics[width=0.5\textwidth]{fig_sxx_temp.pdf}
%     \caption{Noise spectral density $s_{xx}^\text{SQ}(\omega)$ for a single TLF, evaluated at different temperatures. The negative peak is exponentially suppressed as temperature is lowered. Away from the peak, the positive-frequency spectral density is insensitive to the lowering of temperature. When close to the peak, $s_{xx}^\text{SQ}(\omega)$ is enhanced due to the Zeno effect. The inset shows temperature dependence of $s_{xx}^\text{BR}(\omega)$ obtained from the Bloch--Redfield theory. [Parameters used: $k_\text{B}^2 J_0 \Delta^2 / \omega_\text{t}^2 =6.25\,\text{K}^{-2}$, $\omega_\text{D}/k_\text{B}=470\,\textrm{K}$, and $\omega_\text{t}/k_\text{B}=0.08\,\textrm{K}$.]}
%     \label{fig:sxx_temp}
% \end{figure}
%
The second contribution to the TLF noise spectral density is given by $s_{xx}(\omega)$. Following the same procedure as in the previous subsection, but now employing Eqs.~\eqref{eq:sxup} and ~\eqref{eq:sxdown} instead of Eqs.~\eqref{eq:szup} and \eqref{eq:szdown}, we find that the noise spectral density at positive and negative $\omega$ is given by
\begin{equation}\label{eq:sxx}
    s_{xx}(\omega) = \dfrac{4\omega_\text{t}^2 \gamma^+  }{(\omega^2-\omega_\text{t}^2)^2 + \omega^2 (\gamma^+ +\gamma^-)^2},
\end{equation}
where the rates $\gamma^\pm$ are  $\omega$ dependent.
As one can verify, this expression obeys the condition 
\begin{equation}\label{eq:doobeedoo}
s_{xx}(-\omega)= s_{xx}(\omega)e^{-\beta\omega}
\end{equation}
dictated by the fluctuation--dissipation theorem. Again, to distinguish this spectral density from $s^\text{BR}_{xx}(\omega)$ in Sec.~\ref{sec:qrt}, we refer to Eq.~\eqref{eq:sxx} as $s^\text{SQ}_{xx}(\omega)$ in the following. 
In Fig.~\ref{fig:sxx}(a), we show the comparison between $s^\text{SQ}_{xx}(\omega)$ and $s^\text{BR}_{xx}(\omega)$, using the same bath spectral function as before [Eq.~\eqref{eq:jw}]. While $s^\text{BR}_{xx}(\omega)$ consists of two Lorentzians centered at $\pm\omega_\text{t}$, $s^\text{SQ}_{xx}(\omega)$ exhibits additional asymmetry in each of the two peaks, as required by Eq.\ \eqref{eq:doobeedoo}.
The temperature dependence observed for $s_{xx}^\text{SQ}(\omega)$ and $s_{xx}^\text{BR}(\omega)$ [Fig.~\ref{fig:sxx}(b)] differs qualitatively from that of  $s_{zz}(\omega)$. 
The height of the local maximum of $s_{xx}$ close to $\omega\approx -\omega_\text{t}$ is exponentially suppressed as temperature is lowered, according $n_\text{B}(\omega_\text{t})$. At the positive-frequency peak close to $\omega\approx\omega_\text{t}$~\footnote{If the shift $\delta\omega$ of the spectral density’s maxima is small, then a first-order expansion around $\omega=\omega_\text{t}$ can be used to obtain the approximation as $\delta\omega=\frac{\gamma(\omega_\text{t})}{\gamma'(\omega_\text{t})}\big(\frac{2}{\sqrt{4+\gamma'(\omega_\text{t})^2}}-1\big)$}, the opposite occurs: here, the peak height instead increases as temperature is lowered. This is similar to the Purcell effect~\cite{Sete2014} in the near-resonant case, and can be interpreted in terms of the quantum Zeno effect~\cite{Itano1990}: The reduced temperature lowers the decay rate from the TLF to the bath, which results in an increase of the hybridization between the TLF and qubit, and further leads to an enhanced decay rate of the qubit. 
The tail of the peak at $\omega>\omega_\text{t}$ shows only very weak temperature dependence, for reasons analogous to those discussed for $s_{zz}(\omega)$. Namely, based on the  coupling $\kappa\hat{\tau}_x\hat{\sigma}_x$ relevant for $s_{xx}(\omega)$, one finds only one perturbative process involving TLF dephasing and simultaneous qubit relaxation. Since the emitted energy from the qubit is absorbed by the bath, this process is relatively temperature insensitive, thus explaining the weak temperature dependence of $s_{xx}^\text{SQ}(\omega)$.

\subsubsection{Resulting noise spectral density of a single TLF}
The full noise spectral density  is now easily obtained as a linear combination [Eq.~\eqref{eq:spec}] of the longitudinal and transverse contributions $s^\text{SQ}_{zz}(\omega)$ and $s^\text{SQ}_{xx}(\omega)$:
\begin{align}\label{eq:tlf_psd}
\begin{split}
    s(\omega) =& \,\cos^2(\theta) \dfrac{4(p_1^\text{eq}\gamma_\downarrow^++p_0^\text{eq}\gamma_\uparrow^+)}{\omega^2 + (\gamma_\downarrow^++\gamma_\downarrow^- + \gamma_\uparrow^++\gamma_\uparrow^-)^2/4}  \\
    & +\sin^2(\theta)  \dfrac{4\omega_\text{t}^2 \gamma^+  }{(\omega^2-\omega_\text{t}^2)^2 + \omega^2 (\gamma^+ +\gamma^-)^2}.
    \end{split}
\end{align}
This quantity represents the noise from a single TLF. In order to describe charge noise, we consider the combined noise from an ensemble of TLFs with given probability distributions for the TLF parameters $\varepsilon$ and $\Delta$. (These, in turn affect both $\theta=\tan^{-1}(\Delta/\varepsilon)$ and $\omega_\text{t}=\sqrt{\Delta^2+\varepsilon^2}$ in the expression above.) The purpose of the next section is to compute the charge noise $S(\omega)$ from an ensemble average of $s(\omega)$.

\section{Charge-noise Spectral density of an ensemble of Two-Level Fluctuators}\label{sec:ens}

The combined effect of many TLFs can describe some of the experimentally observed properties of charge noise, given an appropriate choice of the underlying distributions for the TLF parameters $\varepsilon,\,\Delta$, and the nature of the TLF--bath coupling strength. 
Borrowing from the approach in Ref.~\onlinecite{phillips1981amorphous}, we model the TLF--bath interaction with the cubic spectral function typical of a phonon bath:
\begin{equation}\label{eq:jwe}
    J(\omega) = J_0 \omega^3 e^{-\omega^2/2\omega_\text{D}^2}.
\end{equation}
Here, we take the coupling parameter $J_0=0.047\,\textrm{ps}^2$ and the Debye frequency $\omega_\text{D}=470$ K, estimated for TLF--phonon interaction in $\textrm{SiO}_2$~\cite{Constantin2009}.
In the usual model of tunneling inside a double-well potential~\cite{phillips1981amorphous}, the tunneling amplitude $\Delta$ is given by $\Delta\propto e^{-\delta}$. Here, $\delta$ primarily depends on the height and width of the barrier between the wells. It is common to assume a uniform distribution for $\delta$~\cite{Phillips1972,Paladino2014}, which results in a log-uniform distribution for $\Delta$.
Distributions used for the bias energy (asymmetry of the double-well potential) $\varepsilon$ vary throughout the literature: A linear distribution~\cite{shnirman2005low} yields the Ohmic spectral density observed for frequencies $\omega > \kb T$ in Ref.~\onlinecite{Astafiev2004}. On the other hand, a uniform distribution~\cite{Constantin2009} is used to reproduce the constant spectral density observed in Ref.~\onlinecite{Martinis2005} for frequencies $\omega/2\pi > 10\,$MHz. To account for both possibilities, we consider the following normalized probability distribution of TLF parameters:
\begin{equation}\label{eq:dist}
    P(\varepsilon,\Delta) = \mathcal{N}(\alpha) \mathcal{B}(\varepsilon,\Delta) \dfrac{\varepsilon^\alpha}{\Delta},
\end{equation}
where $\alpha\in \{0,1\}$ describes the two possible distributions. The boundary function $\mathcal{B}(\varepsilon,\Delta)$ equals unity for $\varepsilon_\text{m}< \varepsilon < \varepsilon_\text{M}$ and $\Delta_\text{m} < \Delta < \Delta_\text{M}$, and vanishes otherwise. The parameter ranges $\varepsilon_\text{m}/\kb = 0$, $\varepsilon_\text{M}/\kb = 4\, \textrm{K}$, $\Delta_\text{m}/\kb = 2\, \mu\textrm{K}$, and $\Delta_\text{M}/\kb = 4\,\textrm{K}$ are taken from Ref.~\onlinecite{Constantin2009}, and the normalization factor is given by
\begin{equation}
    \mathcal{N}(\alpha)^{-1}=
       \bigg(\dfrac{\varepsilon_\textrm{M}+\varepsilon_\textrm{m}}{2}\bigg)^\alpha(\varepsilon_\textrm{M}-\varepsilon_\textrm{m})\ln\bigg( \dfrac{\Delta_\textrm{M}}{\Delta_\textrm{m}}\bigg).
\end{equation}
The spectral density of an ensemble of TLFs is then obtained through
\begin{equation}\label{eq:sbw}
    S(\omega) = N_\textrm{TLF}\iint_{\mathbb{R}^2} \mathrm{d}\varepsilon\, \mathrm{d}\Delta\, P(\varepsilon,\Delta) s(\omega),
\end{equation}
where $s(\omega)$ is the spectral density of a single TLF [Eq.~\eqref{eq:tlf_psd}] and $N_\textrm{TLF}$ is the number of TLFs. For instance, the sample in Ref.~\onlinecite{Constantin2009} with dimensions $400\times40\times800\,\textrm{nm}^3$ and a density of states $n_\textrm{TLF}\approx10^{45}\,\textrm{J}^{-1}\textrm{m}^{-3}$, contains $N_\textrm{TLF}\approx10^3$ TLFs. 
The presence of such TLFs induces
fluctuating charges 
$
    \hat{Q}= (p/L)\hat{\Sigma}_z
$.
Here, $p$ represents the electric dipole moment of a single TLF, and $L$ denotes a sample-dependent characteristic length scale. As a result, $S(\omega)$ is related to charge noise via
$S_\textrm{Q} (\omega)/e^2 = (p/eL)^2 S(\omega)$,
with $p/eL\approx 10^{-4}$ obtained from parameters consistent with Ref.~\onlinecite{Constantin2009}. 

\begin{figure}
    \includegraphics[width=0.48\textwidth]{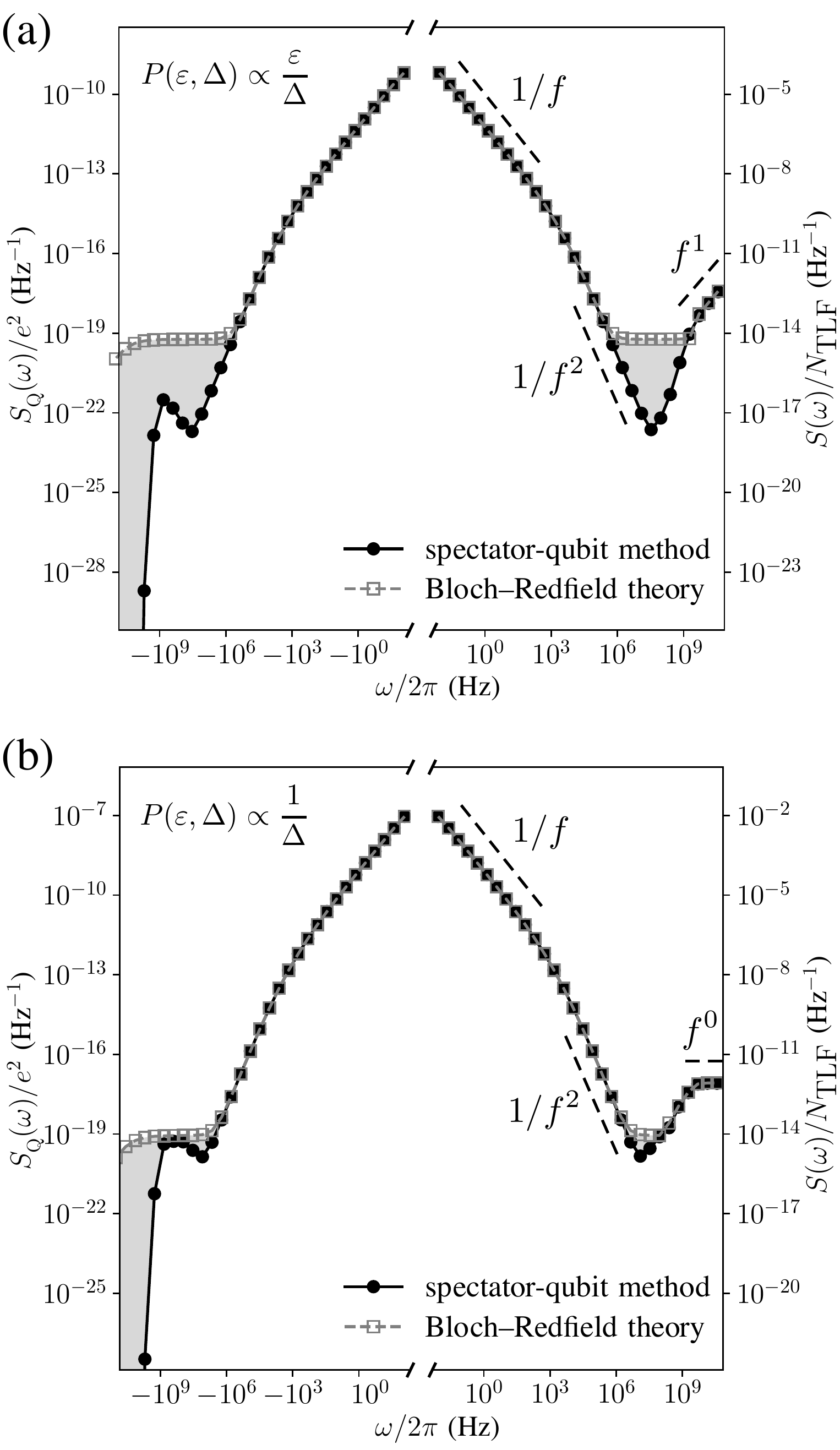}
    \caption{ 
    Charge-noise spectral density of an ensemble of TLFs at 10 mK, with different distributions $P(\varepsilon,\Delta)\propto \varepsilon/\Delta$ in (a) and $P(\varepsilon,\Delta)\propto 1/\Delta$ in (b). In each case, the spectral density is calculated in two ways. The solid black curve represents the results from the spectator-qubit method. For comparison, the gray dashed curve shows $S(\omega)$ computed via the Bloch--Redfield theory. Their difference is highlighted as the shaded region. The dashed lines help to visualize the crossover from $1/f$ to $1/f^2$ in low frequency and the Ohmic (a) or white noise (b) in high frequency. Normalized by the number of TLFs, $S(\omega)/N_\textrm{TLF}$ shown on the right vertical axis supplements the extensive quantity $S_\textrm{Q}(\omega)/e^2$.
    }
    \label{fig:ensemble}
\end{figure}

In Fig.~\ref{fig:ensemble}, we present plots of $S_\text{Q}(\omega)$ for the two choices of probability distributions (i.e., $\alpha=0,\,1$), for a temperature of $10$ mK. The results calculated via the Bloch--Redfield theory are shown for comparison. In the following two subsections we discuss the distinct properties for positive and negative frequencies, respectively.

\subsection{Noise spectral density at positive frequencies}
The noise spectral density at positive frequencies exhibits three regimes with qualitatively different characteristics [see Fig.~\ref{fig:ensemble}(a) and (b)].

(1) At low frequency, we observe a crossover from $1/f$ to $1/f^2$ behavior. 

(2) At high frequency, an Ohmic noise spectrum is obtained for the linear-$\varepsilon$ probability [Fig.~\ref{fig:ensemble}(a)], whereas the spectrum becomes white for the uniform-$\varepsilon$ distribution [Fig.~\ref{fig:ensemble}(b)]. 

(3) An intermediate region exhibiting a local minimum in the noise spectral density connects the low- and high-frequency parts. 
A comparison with the calculation from Sec.~\ref{sec:qrt} shows that the Bloch--Redfield method works well in the low- and high-frequency regimes, but leads to a significantly shallower local minimum in the intermediate region (note the logarithmic scale).

\begin{figure}
    \includegraphics[width=0.5\textwidth]{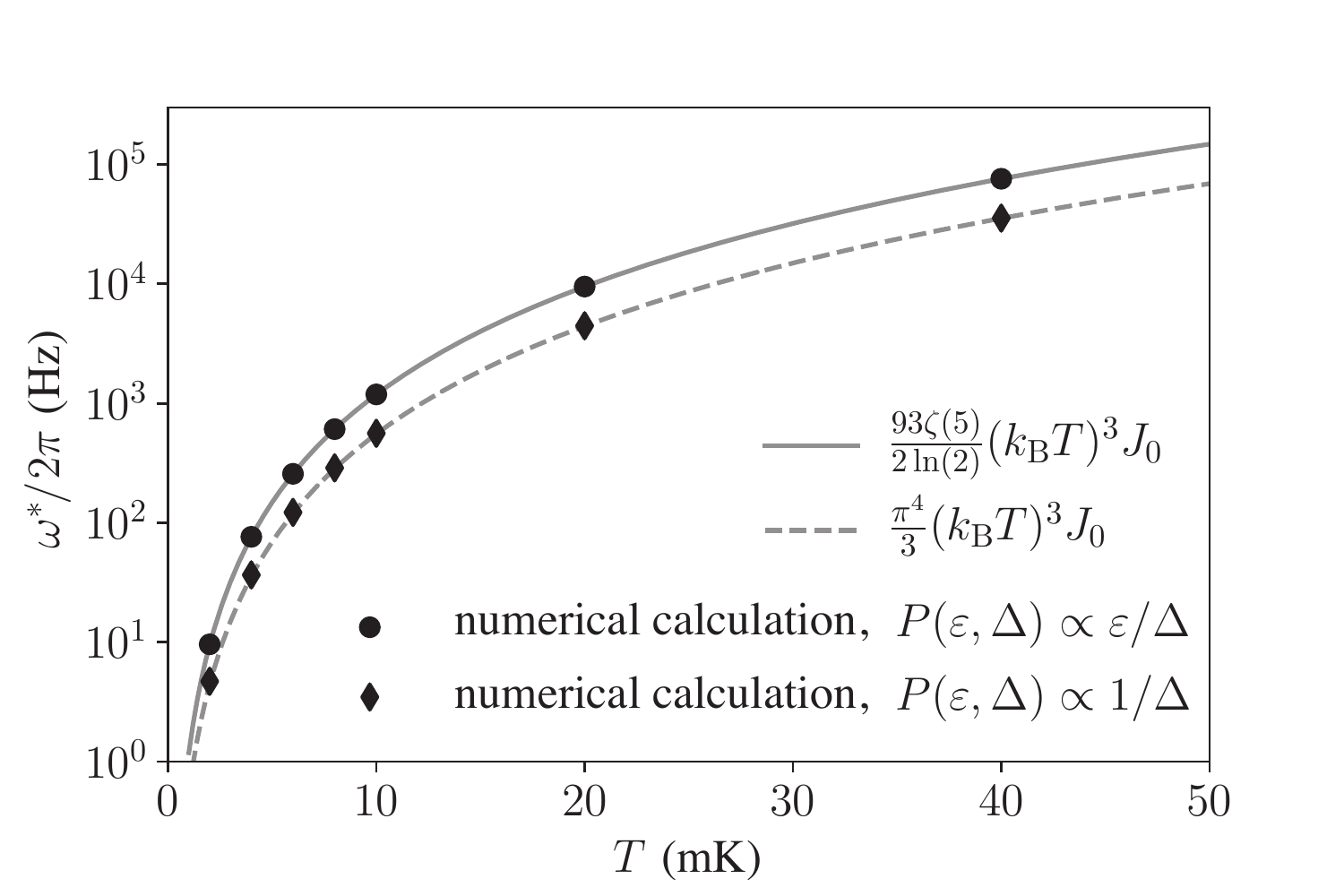}
    \caption{ 
    Temperature dependence of the frequency where the crossover between $1/f$ and $1/f^2$ occurs. Black circles and diamonds are numerical results extracted from the calculations of $S_\textrm{Q}(\omega)$, with $\alpha=1$ and $\alpha=0$, respectively. The analytical expressions for the crossover frequency are in excellent agreement with the data points obtained from numerical calculation.}
    \label{fig:crossover}
\end{figure}

To shed light on the crossover between $1/f$ and $1/f^2$ behavior, we approximate the integral in Eq.~\eqref{eq:sbw} semi-analytically and estimate the crossover frequency $\omega^*$. At low frequencies, $s(\omega)$ is dominated by $s_{zz}(\omega)$, and it is appropriate to use the expression from Eq.~\eqref{eq:sxx_c}, obtained via the Bloch--Redfield theory, as an approximation 
\begin{equation}\label{eq:szz2}
  s(\omega)\approx \cos^2(\theta) s_{zz}(\omega) = \cos^2(\theta) (1-\langle \hat{\sigma}_z \rangle_\text{eq} ^2) \dfrac{2\gamma_1}{\omega^2  + \gamma_1^2}.
\end{equation}
Here, the depolarization rate is given by
\begin{equation}
    \gamma_1 = 2\pi J_0 \omega_\text{t}\Delta^2 \coth\left(\dfrac{\omega_\text{t}}{2k_\text{B}T}\right).
\end{equation}
The average over TLF parameters $\varepsilon$ and $\Delta$, required in Eq.\ \eqref{eq:sbw}, can be converted to an average over $\gamma_1$ and $\omega_\text{t}$, with underlying joint distribution
\begin{equation}\label{eq:dist_gamma}
    P(\gamma_1, \omega_\text{t} ) \propto \dfrac{\omega_\text{t}}{2\gamma_1 }\bigg(\omega_\textrm{t}\sqrt{1-\gamma_1/\gamma_\text{M}}\,\bigg)^{\alpha-1},
\end{equation}
and cutoffs inherited from  $\varepsilon$ and $\Delta$, i.e., $\omega_\textrm{m} = \sqrt{\Delta_\textrm{m}^2+\varepsilon_\textrm{m}^2}$, $\omega_\textrm{M} = \sqrt{\Delta_\textrm{M}^2+\varepsilon_\textrm{M}^2}$, $\gamma_\text{m}(\omega_\textrm{t})=\,2\pi J_0\omega_\text{t}\Delta_\text{m}^2 \coth(\omega_\text{t}/2k_\text{B}T)$, and $\gamma_\text{M}(\omega_\textrm{t})=\,2\pi J_0\omega_\text{t}^3 \coth(\omega_\text{t}/2k_\text{B}T)$.
Taking the average of $s(\omega)$ in Eq.~\eqref{eq:szz2} leads to (see details in Appendix~\ref{app:cross})
\begin{equation*}
    \int_{\omega_\textrm{m}}^{\omega_\textrm{M}} \mathrm{d}\omega_\textrm{t} \int_{\gamma_\text{m}(\omega_\textrm{t})}^{\gamma_\text{M}(\omega_\textrm{t})} \mathrm{d}\gamma_1
    P(\gamma_1, \omega_\text{t} )s(\omega)\propto
     \begin{cases}
       \omega^{-1}, &\omega \ll \omega^* \\ \\
       \omega^{-2}, &\omega \gg \omega^*
    \end{cases},
\end{equation*}
with the crossover frequency
\begin{equation}\label{eq:cross_freq}
    \omega^* \approx
     \begin{cases}
      \dfrac{93\mathcal{\zeta}(5)}{2\ln(2)} (\kb T)^3 J_0 , &P(\varepsilon,\Delta)\propto \varepsilon/\Delta \\ \\
      \dfrac{\pi^4}{3} (\kb T)^3 J_0, &P(\varepsilon,\Delta)\propto 1/\Delta
    \end{cases}.
\end{equation}
We compare this approximation with numerical results as follows. We obtain $S(\omega)$ via numerical integration of Eq.~\eqref{eq:sbw}. A straight line in the log--log scale is generated and extrapolated to larger frequency, by connecting two points of $S(\omega)$ in the $1/f$ regime. Likewise, another straight line connecting points in the $1/f^2$ regime is produced and extrapolated to the lower frequency. The intersection of the two lines is extracted as the crossover frequency $\omega^*$. The above process is repeated for a range of temperature. The crossover frequencies obtained in this way for the linear-$\varepsilon$ and uniform-$\varepsilon$ distributions are shown  in Fig.~\ref{fig:crossover} as circles and diamonds, respectively. Our analytical approximation for the crossover frequency is in excellent agreement with the numerical results.

In the low-frequency and high-frequency regimes, our results match the $1/f$, Ohmic and white-noise behavior discussed in Refs.~\onlinecite{shnirman2005low,Constantin2009}. However, the intermediate region exhibiting a local minimum in the crossover from $1/f$ to white noise was not captured in Ref.~\onlinecite{Constantin2009}. This can be traced back to the use of a fixed Lorentzian linewidth $\gamma_1/2$ [rather than an appropriate distribution in Eq.~\eqref{eq:dist_gamma}] in the calculation of $s_{xx}(\omega)$~\cite{Constantin2009}. Our prediction of a crossover from $1/f$ to $1/f^2$ is consistent with experimental data by Ithier \textit{et al.}~\cite{Ithier2005}, but was not discussed in Refs.~\onlinecite{shnirman2005low,Constantin2009}. A $1/f$ to $1/f^2$ crossover was also mentioned in Ref.~\onlinecite{shnirmanreview}, albeit for a different model including mean-field interactions among TLFs.

\subsection{Noise spectral density at negative frequencies}
Considering the noise spectral density obtained from the spectator-qubit method at negative frequencies, we observe that $S(\omega)$ is approximately symmetric for small $|\omega/\kb T|$, as required by the fluctuation--dissipation theorem. In particular, the crossover from $1/f$ to $1/f^2$ is also visible on the negative-frequency side.
However, for negative frequencies of large magnitude, $|\omega|\gg \kb T$, the spectral density is exponentially suppressed relative to the counterpart on the  positive-frequency side, consistent with the fluctuation--dissipation theorem. 
This exponential suppression, combined with the non-monotonic behavior of $S(\omega)$ in the positive-frequency range, is at the origin of the additional local maximum found in the negative-frequency tail of $S(\omega)$.   

Comparison of the noise spectral density obtained from the spectator-qubit method with the one obtained via the Bloch--Redfield theory shows qualitatively different behavior of the negative-frequency tails. The latter noise spectral density does not exhibit any local extrema for negative frequencies, and the suppression of the negative-frequency tail is much weaker. The latter is related to the aforementioned violation of the fluctuation--dissipation theorem. 

\section{Conclusions}\label{sec:con}
In summary, we have identified violations of the fluctuation--dissipation theorem in the conventional modeling of charge noise, and traced this issue to the missing relevant frequency components of the bath correlation function in the Bloch--Redfield theory. By using the spectator-qubit method (i.e., coupling an auxiliary qubit to the noise source), we recover the relevant frequency components of the bath correlation function, and derive a charge-noise spectral density compatible with the fluctuation--dissipation theorem.
Based on this treatment, we find that $S(\omega)$ exhibits distinct behavior across different frequency ranges: a crossover from $1/f$ to $1/f^2$ at low frequencies, a local minimum at intermediate frequencies, and  Ohmic or white noise behavior at high frequencies.
In line with the fluctuation--dissipation theorem, the negative-frequency part of the spectrum mirrors its positive-frequency counterpart with the necessary exponential suppression factor. 
Our results highlight that the simple model of an ensemble of TLFs generates a noise spectral density with rich behavior in terms of  frequency and temperature dependence. For both we present concrete predictions that can be tested in future experiments on charge noise.

\begin{acknowledgments}
We thank Z.\ Huang, D.\ K.\ Weiss and D.\ G.\ Ferguson for illuminating discussions.
This work was supported by the U.S. Army Research Office under Contracts NO. W911NF-17-C-0024 and NO. W911NF-19-10328 and the Northwestern--Fermilab Center for Applied Physics and Superconducting Technologies.
\end{acknowledgments}

\appendix

\section{Markov approximation in the Schr\"{o}dinger vs.\ interaction picture}\label{app:markov}
The Markov approximation is commonly applied to convert a time-nonlocal master equation into a time-local one. The former involves a time integral of the form $\int_0^t \mathrm{d}s\, F[\hat{\rho}(s)]$.  Assuming that the system density matrix $\hat{\rho}$ undergoes dynamics that is slow compared to the fast equilibration of the bath, one may approximate that integral by replacing $\hat{\rho}(s)\to \hat{\rho}(t)$, i.e., the system density matrix at the present time. This Markov approximation can be applied within either the Schr\"odinger or the interaction picture, and each choice generally leads to a different master equation and a corresponding system evolution. Depending on the specific dynamics (e.g., oscillation mode, relaxation mode, etc.), one choice may be more appropriate than the other. To illustrate this point, we discuss two representative examples in the following.

\subsection{Dephasing dynamics: Markov approximation in the interaction picture}
In the first case, we consider a two-level system coupled longitudinally to a thermal bath. After performing the Markov approximation in either of the two pictures, the evolution of the density matrix in the Schr\"odinger picture follows:
\begin{equation}\label{eq:a1}
    \hat{\rho}(t)=
    \left(
        \begin{array}{cc}
             \rho_\text{ee}(0) &  \rho_\text{eg}(0)e^{-i\omega_0 t - \gamma t} \\
              \rho_\text{ge}(0)e^{i\omega_0 t- \gamma t} & \rho_\text{gg}(0) \\
        \end{array}
    \right).
\end{equation}
Here, $\omega_0$ is the frequency of the two-level system, and $\gamma$ denotes the dephasing rate. (We note that the value of $\gamma$ will generally  depend on whether the Markov approximation is applied inside the Schr\"odinger picture or the interaction picture.)
Eq.~\eqref{eq:a1} describes dephasing dynamics, with diagonal elements remaining constant, but off-diagonal elements undergoing exponentially damped oscillations. When transformed into the interaction picture (denoted by a tilde), the same evolution takes on the form
\begin{equation}\label{eq:a2}
    \Tilde{\hat{\rho}}(t)=
    \left(
        \begin{array}{cc}
             \rho_\text{ee}(0) &  \rho_\text{eg}(0)e^{- \gamma t} \\
              \rho_\text{ge}(0)e^{- \gamma t} & \rho_\text{gg}(0) \\
        \end{array}
    \right).
\end{equation}
As opposed to Eq.~\eqref{eq:a1}, the interaction-picture evolution does not show any oscillatory behavior for the off-diagonal elements. Given that $\gamma \ll \omega_0$, this implies that the dynamics is significantly faster in the Schr\"odinger picture as compared to the interaction picture. As a result, the evolution described by Eq.~\eqref{eq:a2} is more suitable for the slow-dynamics assumption that underlies the Markov approximation. Hence, in this example, one should expect to obtain more accurate results when employing the Markov approximation in the interaction picture.

\subsection{Relaxation dynamics: Markov approximation in the Schr\"odinger picture}
For the second example, we consider the setup described in Sec.~\ref{subsec:qubitandtlf}, where a thermal bath is transversely coupled to a TLF, which in turn couples to an auxiliary qubit. Here, we are particularly interested in the relaxation dynamics of the qubit. After performing the Markov approximation in either of the two pictures, the reduced density matrix of the qubit undergoes relaxation dynamics. In the Schr\"{o}dinger picture this takes the form 
\begin{equation}\label{eq:a3}
    \hat{\rho}(t)=\,\hat{\rho}_\textrm{eq} + 
    \left(
        \begin{array}{cc}
             a &  b \\
             b^* & -a \\
        \end{array}
    \right)
    e^{-\gamma t},
\end{equation}
where $\hat{\rho}_\textrm{eq}$ denotes the equilibrium density matrix, $a$ and $b$ are constants forming a traceless coefficient matrix, and $\gamma$ is the relaxation rate. Transformed into the interaction picture, the same evolution is described by
\begin{equation}\label{eq:a4}
    \tilde{\hat{\rho}}(t)=\,\hat{\rho}_\textrm{eq} + 
    \left(
        \begin{array}{cc}
             a &  b\,e^{ i\omega_\textrm{q}t} \\
             b^*\,e^{-i\omega_\textrm{q}t} & -a \\
        \end{array}
    \right)e^{-\gamma t},
\end{equation}
where $\omega_\textrm{q}$ is the qubit frequency. Contrasting the latter expression with Eq.\ \eqref{eq:a3} reveals that the absence of oscillations in the Schr\"odinger picture renders the dynamics slow compared to the interaction picture.  As a result, the Markov approximation is here more appropriate within the Schr\"{o}dinger picture for a more accurate description of the qubit's relaxation dynamics. 

\section{Expression for the evolution superoperator $\Lambda_0$}\label{app:evo_mat}
The evolution superoperator $\Lambda_0$ [Eq.\ \eqref{eq:lambda_evo}] becomes block diagonal when expressed in the basis formed by the eigenstates of the combined qubit--TLF system:
\begin{align*}
    \Lambda_0=& \,\textrm{diag}(M_1,\, M_1,\, M_2 - i \omega_\text{q},\, M_2 + i \omega_\text{q},\, 
    M_3 - i \omega_\text{t} \sigma_z,\\ 
    & M_3 - i \omega_\text{t} \sigma_z,\,M_4 - i \omega_\text{q} - i \omega_\text{t} \sigma_z,\, M_4 + i \omega_\text{q} - i \omega_\text{t} \sigma_z),
\end{align*}
with the definitions 
\begin{equation*}
    M_1=
    \left(
        \begin{array}{cc}
             -\gamma_\uparrow & \quad \phantom{-}\gamma_\downarrow \\
              \phantom{-}\gamma_\uparrow &\quad -\gamma_\downarrow \\
        \end{array}
    \right),
\end{equation*}
\begin{equation*}
    M_2=\dfrac{1}{2}
    \left(
        \begin{array}{cc}
             -\gamma_\uparrow^- - \gamma_\uparrow^+ &\quad -\gamma_\downarrow^- - \gamma_\downarrow^+ \\
              -\gamma_\uparrow^- - \gamma_\uparrow^+ &\quad  -\gamma_\downarrow^- - \gamma_\downarrow^+ \\
        \end{array}
    \right),
\end{equation*}
\begin{equation*}
    M_3=-\gamma(0)
    \left(
        \begin{array}{cc}
             \phantom{-}1 &\quad  -1 \\
              -1 &\quad \phantom{-}1\\
        \end{array}
    \right),
\end{equation*}
\begin{equation*}
    M_4=-\dfrac{\gamma^- + \gamma^+}{2}
    \left(
        \begin{array}{cc}
             1 &\quad 1 \\
             1 &\quad 1 \\
        \end{array}
    \right).
\end{equation*}

\section{Derivation of qubit relaxation and excitation rates in Eq.~\eqref{eq:depolarization} from degenerate perturbation theory}\label{app:dept}
Here we derive the qubit relaxation and excitation rates induced by the coupling to the noise source (TLF and bath). Treating the coupling between qubit and TLF perturbatively, we expand the evolution matrix in Eq.~\eqref{eq:evo_mat} in powers of $\kappa$
\begin{equation}\label{eq:c1}
    \Lambda = \Lambda_0 + \kappa \Lambda_1 + \kappa^2 \Lambda_2 + \mathcal{O}(\kappa^3).
\end{equation} 
When the coupling is absent (i.e., $\kappa=0$), each of the two eigenvectors of $\Lambda_0$ with eigenvalue zero [Eqs.~\eqref{eq:zero_1} and \eqref{eq:zero_2}] is a product state with the TLF in equilibrium and the qubit occupying either the ground or the excited state. In the presence of the coupling, the twofold degeneracy is lifted, which results in one stationary state and one mode describing the qubit depolarization.
The relevant dynamics is governed by
\begin{equation}
    \dfrac{\mathrm{d}}{\mathrm{d}t} | \rho (t) )= \Lambda | \rho (t)).
\end{equation}

In general, the solution to the above equation for an initial state $|\rho(0))$ has the form of 
\begin{equation}
    | \rho (t)) = \sum_j | \varrho_j)(\varphi_j|\rho(0))e^{\chi_j t},
\end{equation}
where $| \varrho_j)$ and $(\varphi_j|$ are the right and left eigenvectors of non-Hermitian matrix $\Lambda$ with eigenvalue $\chi_j$.
To extract the relaxation rate, we initialize the qubit in the excited state (with TLF in equilibrium), and monitor the population increase of the ground state:
\begin{equation}\label{eq:c44}
    (\phi_\textrm{g}| \rho (t)) = \sum_j (\phi_\textrm{g}| \varrho_j)(\varphi_j|\rho_\textrm{e})e^{\chi_j t}.
\end{equation}
In the following, we use index $j=0, 1$ to denote the zero mode and depolarization mode of $\Lambda$, which reduce to the two zero modes of $\Lambda_0$ as $\kappa\to 0$. It is expected that the amplitude $(\phi_\textrm{g}| \varrho_j)(\varphi_j|\rho_\textrm{e})$ is of order $\kappa^0$ for $j=0,1$, and of order $\kappa^2$ for $j\ge2$, respectively. Hence, Eq.~\eqref{eq:c44} can be approximated by
\begin{equation}
    (\phi_\textrm{g}| \rho (t)) \approx \sum_{j=0,1} (\phi_\textrm{g}| \varrho_j)(\varphi_j|\rho_\textrm{e})e^{\chi_j t}.
\end{equation}
Expanding this population for times $t$ small compared to the depolarization time $|\chi_1|^{-1}$, we obtain
\begin{equation}
    \dfrac{\mathrm{d}}{\mathrm{d}t}(\phi_\textrm{g}| \rho (t)) \approx (\phi_\textrm{g}| \varrho_1)(\varphi_1|\rho_\textrm{e})\chi_1. 
\end{equation}
Hence, we identify the relaxation rate as
\begin{equation}\label{eq:cc5}
    \Gamma_\downarrow \approx (\phi_\textrm{g}| \varrho_1)(\varphi_1|\rho_\textrm{e})\chi_1.
\end{equation}
Similarly, the excitation rate is obtained by initializing the system in a state with the qubit in the ground state:
\begin{equation}\label{eq:cc6}
    \Gamma_\uparrow \approx (\phi_\textrm{e}| \varrho_1)(\varphi_1|\rho_\textrm{g})\chi_1.
\end{equation}
To further evaluate these expressions, it is necessary to diagonalize $\Lambda$ in the degenerate subspace of $\Lambda_0$. In the following, we calculate the eigenvalues and eigenvectors of $\Lambda$ up to second order in $\kappa$ using degenerate perturbation theory. 

We start from the eigenvalue equation 
\begin{equation}\label{eq:c2}
    \Lambda|\varrho_j)=\chi_j |\varrho_j).
\end{equation}
Similar to Eq.~\eqref{eq:c1}, we expand eigenvectors and eigenvalues of $\Lambda$ up to second order in $\kappa$:
\begin{align}
    \chi_j =& \,\chi_j^{(0)} + \kappa\chi_j^{(1)} +\kappa^2 \chi_j^{(2)} + \mathcal{O}(\kappa^3),\label{eq:c3}\\
    |\varrho_j) =& \,|\varrho_j^{(0)}) + \kappa |\varrho_j^{(1)}) + \kappa^2 |\varrho_j^{(2)}) + \mathcal{O}(\kappa^3),\label{eq:c4}
\end{align}
where $|\varrho_j^{(0)})$ and $\chi_j^{(0)}$ are the eigenvectors and eigenvalues of $\Lambda_0$. For diagonalization in the degenerate subspace, it is convenient to decompose the vector space $\mathcal{V}$ spanned by all right eigenvectors of $\Lambda_0$ into a direct sum of the degenerate subspace $\mathcal{D}$ and its complementary subspace $\overline{\mathcal{D}}$, such that $\mathcal{V}=\mathcal{D}\oplus\overline{\mathcal{D}}$.
Similarly, we define $\mathcal{W}=\mathcal{B}\oplus\overline{\mathcal{B}}$ as the vector space spanned by left eigenvectors of $\Lambda_0$, along with $\mathcal{B}$ the degenerate subspace and its complementary subspace $\overline{\mathcal{B}}$. 
The notation for eigenvalues and eigenvectors of $\Lambda_0$ is summarized in Table~\ref{tab:tab1}.
\begin{table}[b]
\caption{\label{tab:tab1}
Notation for eigenvalues and eigenvectors of $\Lambda_0$.}
\begin{ruledtabular}
\begin{tabular}{llll}
 &
\begin{tabular}[c]{@{}c@{}}eigenvalue \\of $\Lambda_0$\end{tabular} &
\begin{tabular}[c]{@{}c@{}}eigenvector \\of $\Lambda_0$\end{tabular} &
\begin{tabular}[c]{@{}c@{}}eigenvector \\of $\Lambda_0^\dag$\end{tabular} \\
\colrule
\begin{tabular}[c]{@{}c@{}}degenerate \\ subspaces \\$\mathcal{D}$, $\mathcal{B}$ \\[5pt]complem. \\ subspaces \\$\overline{\mathcal{D}}$, $\overline{\mathcal{B}}$ \end{tabular}  &  $\begin{rcases*} \lambda_\alpha \\ \\[6pt] \overline{\lambda}_\mu \end{rcases*} \chi_j^{(0)} $ &  $\begin{rcases*} |\rho_\alpha ) \\ \\[6pt] |\overline{\rho}_\mu ) \end{rcases*} \big| \varrho_j^{(0)}\big) $ &  $\begin{rcases*} (\phi_\beta| \\ \\[6pt] (\overline{\phi}_\nu| \end{rcases*} \big(\varphi_j^{(0)}\big| $\\
\end{tabular}
\end{ruledtabular}
\end{table}
The eigenvectors $|\varrho_j)$ of $\Lambda$ can thus be expanded in the eigenbasis of $\Lambda_0$ as follows:
\begin{equation}\label{eq:c5}
    |\varrho_j)  =\sum_\alpha c_{j\alpha} |\rho_\alpha ) + \sum_\mu d_{j \mu} |\overline{\rho}_\mu)+ \mathcal{O}(\kappa^3),
\end{equation}
with $|\rho_\alpha)\in\mathcal{D}$, and $|\overline{\rho}_\mu)\in\overline{\mathcal{D}}$. In the following, we focus on the eigenvectors which reduce to the two zero modes of $\Lambda_0$, i.e., we consider $j=0,1$.
Comparing with Eq.~\eqref{eq:c4}, it follows that the coefficients $ c_{j\alpha}$ and $d_{j\mu}$ are of order $\kappa^0$ and $\kappa^1$, respectively~\footnote{\label{f3}For $j\neq 0,1$, similar observation shows that $ c_{j\alpha}$ and $d_{j\mu}$ are of order $\kappa^1$ and $\kappa^0$, respectively.}.
Substituting $|\varrho_j)$ [Eq.~\eqref{eq:c5}] and $\chi_j$ [Eq.~\eqref{eq:c3}] into the eigenvalue equation [Eq.~\eqref{eq:c2}] yields 
\begin{align}
    \begin{split}\label{eq:c11}
    &\sum_\alpha c_{j\alpha} ( -\kappa\chi_j^{(1)} - \kappa^2\chi_j^{(2)} + \kappa \Lambda_1 + \kappa^2 \Lambda_2)|\rho_\alpha)  \\
    &+ \sum_\mu d_{j\mu} (\overline{\lambda}_\mu -\kappa\chi_j^{(1)} + \kappa \Lambda_1 )|\overline{\rho}_\mu) + \mathcal{O}(\kappa^3) = 0,
    \end{split}
\end{align}
where $\overline{\lambda}_\mu$ is the eigenvalue associated with $|\overline{\rho}_\mu)$.
Projecting \eqref{eq:c11} onto the states $( \phi_\beta |\in\mathcal{B}$ and $( \overline{\phi}_\nu |\in\overline{\mathcal{B}}$ yields
\begin{align}
    \label{eq:ca}
    \begin{split}
        \kappa \sum_\mu d_{j\mu} ( \phi_\beta | \Lambda_1 | \overline{\rho}_\mu) + \kappa^2 \sum_\alpha c_{j\alpha} ( \phi_\beta | \Lambda_2 | \rho_\alpha ) \\
        =  c_{j\beta} (\kappa\chi_j^{(1)} + \kappa^2 \chi_j^{(2)}),
    \end{split}\\
    \label{eq:dnu}
    \begin{split}
        \kappa \sum_\alpha c_{j\alpha} ( \overline{\phi}_\nu | \Lambda_1 | \rho_\alpha ) + 
        \kappa \sum_\mu d_{j\mu} ( \overline{\phi}_\nu | \Lambda_1 | \overline{\rho}_\mu )\\
        = d_{j\nu} (-\kappa\chi_j^{(1)} - \overline{\lambda}_\nu), 
    \end{split}
\end{align}
which holds up to (and including) order $\kappa^2$.
Note that the terms proportional to $( \phi_\beta | \Lambda_1 | \rho_\alpha)$ vanish, since $\Lambda_1$ is offdiagonal in the qubit subspace. Moreover, Eq.~\eqref{eq:ca} shows that $\chi_j^{(1)}$ is zero, by comparing orders of $\kappa$ and recalling that $d_{j\mu}\sim\mathcal{O}(\kappa)$. Since eigenvectors in the degenerate subspace are only associated with $c_{j\alpha}$, we proceed as follows. Solving Eq.\ \eqref{eq:dnu} for $d_{j\mu}$ results in an expression in terms of  $c_{j\alpha}$ which can then be substituted into Eq.\ \eqref{eq:ca}. Since the term involving $d_{j\mu}$ in \eqref{eq:ca} carries a factor of $\kappa$, it is sufficient to retain only $\mathcal{O}(\kappa)$ terms for $d_{j\mu}$, which yields
\begin{equation}\label{eq:cx5}
    d_{j\nu} =  -\dfrac{\kappa}{\overline{\lambda}_\nu}\sum_\alpha c_{j\alpha} ( \overline{\phi}_\nu | \Lambda_1 | \rho_\alpha ).
\end{equation}
Upon substitution back  into Eq.~\eqref{eq:ca}, we find
\begin{align*}
    c_{j\beta} \chi_j^{(2)} =& \,  \sum_\alpha c_{j\alpha} ( \phi_\beta | \Lambda_2 | \rho_\alpha ) \nonumber\\
    & -  \sum_\alpha c_{j\alpha}\sum_\mu \overline{\lambda}_\mu^{-1}( \phi_\beta | \Lambda_1 | \overline{\rho}_\mu )( \overline{\phi}_\mu | \Lambda_1 | \rho_\alpha ) .
\end{align*}
Note that this is an eigenvalue equation for $\chi_j^{(2)}$ involving the matrix $\Lambda_\textrm{m}$ defined by 
\begin{equation}\label{eq:c17}
    \Lambda_\textrm{m} = P\Lambda_2 P
    -P\Lambda_1 (\mathrm{1}-P)\Lambda_0^{-1} (\mathrm{1}-P) \Lambda_1 P,
\end{equation}
where $P$ is the projector onto the degenerate subspace [Eq.~\eqref{eq:proj}].
Thus, we obtain $\chi_j^{(2)}$ and $\{c_{j\beta}\}$ by solving the above eigenvalue equation. Since both $\chi_j^{(0)}$ and $\chi_j^{(1)}$ are zero ($j=0,1$), $\kappa^2\chi_j^{(2)}$ approximates the eigenvalue $\chi_j$ of $\Lambda$ up to $\mathcal{O}(\kappa^2)$. Moreover, $\{c_{j\beta}\}$ determines the approximate eigenvectors of $\Lambda$ in the degenerate subspace of $\Lambda_0$ [see Eq.~\eqref{eq:c5}]. Plugging the approximate eigenvalue and eigenvector into Eqs.~\eqref{eq:cc5} and \eqref{eq:cc6} yields the relaxation and excitation rates. Note that a compact form of these rates can be derived by employing the eigendecomposition of $\Lambda_\textrm{m}$:
\begin{equation}
    \kappa^2\Lambda_\textrm{m} \approx | \varrho_0)(\varphi_0|\chi_0 + | \varrho_1)(\varphi_1|\chi_1.
\end{equation}
Using the fact that one eigenvalue of $\Lambda_\textrm{m}$ is zero yields
\begin{equation}
    \kappa^2(\phi_\textrm{e}| \Lambda_\textrm{m} |\rho_\textrm{g})
    \approx
    (\phi_\textrm{e}| \varrho_1)(\varphi_1|\rho_\textrm{g})\chi_1 .
\end{equation}
Therefore, the excitation and relaxation rates in Eqs.~\eqref{eq:cc5} and \eqref{eq:cc6} can be rewritten as
\begin{equation}
    \Gamma_\downarrow \approx \kappa^2(\phi_\text{g}|\Lambda_\textrm{m}|\rho_\text{e} ) , \qquad
    \Gamma_\uparrow \approx \kappa^2(\phi_\text{e}|\Lambda_\textrm{m}|\rho_\text{g} ).
\end{equation}

\section{Derivation of the crossover frequency in Eq.~\eqref{eq:cross_freq}}\label{app:cross}
\subsection{Case: linear distribution in $\varepsilon$}
The low-frequency part of $S(\omega)$ is dominated by $s_{zz}(\omega)$, and can be approximated by
\begin{equation*}
    S(\omega)  \approx \int_{\omega_\textrm{m}}^{\omega_\textrm{M}} \mathrm{d}\omega_\textrm{t}  \int_{\gamma_\text{m}(\omega_\textrm{t})}^{\gamma_\text{M}(\omega_\textrm{t})} \mathrm{d}\gamma_1 P(\gamma_1, \omega_\text{t} ) s_{zz}(\omega) \cos^2(\theta).
\end{equation*}
The joint probability distribution is 
\begin{equation}
     P(\gamma_1, \omega_\text{t} )=  \mathcal{N}_1 \dfrac{\omega_\text{t}}{2\gamma_1 },
\end{equation}
where $\mathcal{N}_1$ is a normalization factor.
At low frequency, it is appropriate to use the expression from Eq.~\eqref{eq:sxx_c} for $s_{zz}(\omega)$, obtained via the Bloch--Redfield theory, as an approximation: 
\begin{equation}
    s_{zz}(\omega) \approx (1-\langle \hat{\sigma}_z \rangle_\text{eq}^2)\dfrac{2\gamma_1}{\omega^2  + \gamma_1^2},
\end{equation}
where $\langle \hat{\sigma}_z \rangle_\text{eq}=\tanh(\omega_\textrm{t}/2\kb T)$.
After converting
\begin{equation}
    \cos^2(\theta) = 1-\dfrac{\Delta^2}{\omega_\textrm{t}^2} = 1 - \dfrac{\gamma_1}{\gamma_\text{M}},
\end{equation}
we obtain $S(\omega)$ as follows:
\begin{align*}
    S(\omega) =& \,\int_{\omega_\textrm{m}}^{\omega_\textrm{M}} \mathrm{d}\omega_\textrm{t}  \int_{\gamma_\text{m}(\omega_\textrm{t})}^{\gamma_\text{M}(\omega_\textrm{t})} \mathrm{d}\gamma_1 
    \,\mathcal{N}_1 \omega_\text{t}\dfrac{1-\langle \hat{\sigma}_z \rangle_\text{eq}^2}{\omega^2  + \gamma_1^2} \bigg(1 - \dfrac{\gamma_1}{\gamma_\text{M}}\bigg) \\
    =& \, \mathcal{N}_1 \int_{\omega_\textrm{m}}^{\omega_\textrm{M}} \mathrm{d}\omega_\textrm{t}\, \omega_\textrm{t} \bigg\{ 
    \dfrac{1}{\omega}\bigg[\tan^{-1}\left(\dfrac{\gamma_\textrm{M}}{\omega}\right) - \tan^{-1}\left( \dfrac{\gamma_\textrm{m}}{\omega}\right)\bigg]\\ 
    &- \dfrac{1}{2\gamma_\textrm{M}} \ln (\dfrac{\gamma_\textrm{M}^2 + \omega^2}{\gamma_\textrm{m}^2 + \omega^2})
    \bigg\}(1-\langle \hat{\sigma}_z \rangle_\text{eq}^2).
\end{align*}
For $\gamma_\textrm{m} \ll \omega \ll \gamma_\textrm{M}$, the above expression simplifies to
\begin{equation}
    S(\omega) 
    \approx  \,\mathcal{N}_1\int_{\omega_\textrm{m}}^{\omega_\textrm{M}} \mathrm{d}\omega_\textrm{t}\, \omega_\textrm{t} \, \dfrac{1}{\omega}  \dfrac{\pi}{2}(1-\langle \hat{\sigma}_z \rangle_\text{eq}^2).
\end{equation}
For $\omega \gg \gamma_\textrm{M}$, we obtain
\begin{equation}
    S(\omega) 
    \approx \, \mathcal{N}_1\int_{\omega_\textrm{m}}^{\omega_\textrm{M}} \mathrm{d}\omega_\textrm{t}\, \omega_\textrm{t} \, \dfrac{\gamma_\textrm{M}}{2\omega^2}(1-\langle \hat{\sigma}_z \rangle_\text{eq}^2).
\end{equation}
Note that both $\gamma_\textrm{M}$ and $\gamma_\textrm{m}$ depend on $\omega_\textrm{t}$:
\begin{align}
    \gamma_\text{m}(\omega_\textrm{t})=& \,2\pi J_0\omega_\text{t}\Delta_\text{m}^2 \coth\left(\dfrac{\omega_\text{t}}{2k_\text{B}T}\right), \label{eq:gmin}\\
    \gamma_\text{M}(\omega_\textrm{t})=& \,2\pi J_0\omega_\text{t}^3 \coth\left(\dfrac{\omega_\text{t}}{2k_\text{B}T}\right).\label{eq:gmax}
\end{align}
In short the low-frequency part of $S(\omega)$ calculated using linear-$\varepsilon$ distribution can be approximated by
\begin{equation*}
    S(\omega) \approx \,\mathcal{N}_1\int_{\omega_\textrm{m}}^{\omega_\textrm{M}} \mathrm{d}\omega_\textrm{t}\, \omega_\textrm{t} (1-\langle \hat{\sigma}_z \rangle_\text{eq}^2)\cdot
     \begin{cases}
      \dfrac{\pi}{2\omega} , &\gamma_\text{m}\ll\omega \ll \gamma_\text{M} \\ \\
       \dfrac{\gamma_\textrm{M}}{2\omega^2}, &\omega \gg \gamma_\text{M}
    \end{cases}.
\end{equation*}
Now we evaluate the integration over $\omega_\textrm{t}$. When $\omega$ is sufficiently small such that the $1/f$ noise is dominant, the integral can be approximated by 
\begin{align}
    S(\omega)_{1/f} \approx & \,\mathcal{N}_1\int_0^\infty \mathrm{d}\omega_\textrm{t}\, \omega_\textrm{t} \bigg[1-\tanh^2\bigg(\dfrac{\omega_\textrm{t}}{2\kb T}\bigg)\bigg]\dfrac{\pi}{2\omega} \nonumber\\
    =& \,\mathcal{N}_1\dfrac{\pi}{2\omega} (2\kb T)^2\ln(2).
\end{align}
When $\omega$ is large enough such that the $1/f^2$ noise is dominant, the integral can be approximated by 
\begin{align}
    S(\omega)_{1/f^2} \approx & \,\mathcal{N}_1\int_0^\infty \mathrm{d}\omega_\textrm{t}\, \omega_\textrm{t}  \bigg[1-\tanh^2\bigg(\dfrac{\omega_\textrm{t}}{2\kb T}\bigg)\bigg]\dfrac{\gamma_\textrm{M}}{2\omega^2} \nonumber\\
    =& \,\mathcal{N}_1\dfrac{2\pi J_0}{\omega^2} (2\kb T)^5\dfrac{93}{64}\mathcal{\zeta}(5).
\end{align}
The crossover frequency of $S(\omega)_{1/f}$ and $S(\omega)_{1/f^2}$ is then identified as 
\begin{equation}
    \omega^* = \dfrac{93\mathcal{\zeta}(5)}{2\ln(2)} (\kb T)^3J_0.
\end{equation}

\subsection{Case : uniform distribution in $\varepsilon$}
In this case, the joint probability distribution is
\begin{equation}
    P(\gamma_1, \omega_\text{t} )=  \mathcal{N}_0 \dfrac{1}{2\gamma_1} \bigg(1-\dfrac{\gamma_1}{\gamma_\textrm{M}}\bigg)^{-1/2},
\end{equation}
where $\mathcal{N}_0$ is the normalization factor. Similar argument leads to the following approximated $S(\omega)$: 
\begin{align*}
    S(\omega) \approx& \int_{\omega_\textrm{m}}^{\omega_\textrm{M}} \mathrm{d}\omega_\textrm{t}  \int_{\gamma_\text{m}(\omega_\textrm{t})}^{\gamma_\text{M}(\omega_\textrm{t})} \mathrm{d}\gamma_1 
    \,\mathcal{N}_0 \dfrac{1-\langle \hat{\sigma}_z \rangle_\text{eq}^2}{\omega^2  + \gamma_1^2} \sqrt{1 - \dfrac{\gamma_1}{\gamma_\text{M}}} \\
    =& \,\mathcal{N}_0 \int_{\omega_\textrm{m}}^{\omega_\textrm{M}} \mathrm{d}\omega_\textrm{t}\, (1-\langle \hat{\sigma}_z \rangle_\text{eq}^2)\\
    & \times\dfrac{2}{\omega}\Im\left[ \sqrt{1+\dfrac{i\omega}{\gamma_\textrm{M}}} \tan^{-1}\left(\sqrt{\dfrac{\gamma_\textrm{M}-\gamma_\textrm{m}}{\gamma_\textrm{M}+i\omega}}\,\right)\right] .
\end{align*}
For $\gamma_\textrm{m} \ll \omega \ll \gamma_\textrm{M}$, the above expression simplifies to
\begin{equation}
    S(\omega) 
    \approx  \,\mathcal{N}_0\int_{\omega_\textrm{m}}^{\omega_\textrm{M}} \mathrm{d}\omega_\textrm{t} \, \dfrac{1}{\omega}  \dfrac{\pi}{2}(1-\langle \hat{\sigma}_z \rangle_\text{eq}^2).
\end{equation}
For $\omega \gg \gamma_\textrm{M}$, we obtain
\begin{equation}
    S(\omega) 
    \approx  \,\mathcal{N}_0\int_{\omega_\textrm{m}}^{\omega_\textrm{M}} \mathrm{d}\omega_\textrm{t}\, \dfrac{2\gamma_\textrm{M}}{3\omega^2}(1-\langle \hat{\sigma}_z \rangle_\text{eq}^2).
\end{equation}
In short the low-frequency part of $S(\omega)$ calculated using constant-$\varepsilon$ distribution can be approximated by
\begin{equation*}
    S(\omega) \approx \,\mathcal{N}_0\int_{\omega_\textrm{m}}^{\omega_\textrm{M}} \mathrm{d}\omega_\textrm{t} \,(1-\langle \hat{\sigma}_z \rangle_\text{eq}^2)\, \cdot
     \begin{cases}
      \dfrac{\pi}{2\omega} , &\gamma_\text{m}\ll\omega \ll \gamma_\text{M} \\ \\
       \dfrac{2\gamma_\textrm{M}}{3\omega^2}, &\omega \gg \gamma_\text{M}
    \end{cases}.
\end{equation*}
Now we evaluate the integration over $\omega_\textrm{t}$. When $\omega$ is sufficiently small such that the $1/f$ noise is dominant, the integral can be approximated by 
\begin{align}
    S(\omega)_{1/f} \approx & \,\mathcal{N}_0\int_0^\infty \mathrm{d}\omega_\textrm{t}\,  \bigg[1-\tanh^2\bigg(\dfrac{\omega_\textrm{t}}{2\kb T}\bigg)\bigg]\dfrac{\pi}{2\omega}\nonumber\\
    =& \,\mathcal{N}_0\dfrac{\pi}{2\omega} 2\kb T.
\end{align}
When $\omega$ is large enough such that the $1/f^2$ noise is dominant, the integral can be approximated by 
\begin{align}
    S(\omega)_{1/f^2} \approx & \,\mathcal{N}_0\int_0^\infty \mathrm{d}\omega_\textrm{t}\,  \bigg[1-\tanh^2\bigg(\dfrac{\omega_\textrm{t}}{2\kb T}\bigg)\bigg]\dfrac{2\gamma_\textrm{M}}{3\omega^2}\nonumber\\
    =& \,\mathcal{N}_0\dfrac{2\pi J_0}{\omega^2} (2\kb T)^4\dfrac{\pi^4}{96}.
\end{align}
The crossover frequency of $S(\omega)_{1/f}$ and $S(\omega)_{1/f^2}$ is then identified as 
\begin{equation}
    \omega^* = \dfrac{\pi^4}{3} (\kb T)^3 J_0.
\end{equation}

\bibliography{main.bib}

\begin{thebibliography}{54}%
\makeatletter
\providecommand \@ifxundefined [1]{%
 \@ifx{#1\undefined}
}%
\providecommand \@ifnum [1]{%
 \ifnum #1\expandafter \@firstoftwo
 \else \expandafter \@secondoftwo
 \fi
}%
\providecommand \@ifx [1]{%
 \ifx #1\expandafter \@firstoftwo
 \else \expandafter \@secondoftwo
 \fi
}%
\providecommand \natexlab [1]{#1}%
\providecommand \enquote  [1]{``#1''}%
\providecommand \bibnamefont  [1]{#1}%
\providecommand \bibfnamefont [1]{#1}%
\providecommand \citenamefont [1]{#1}%
\providecommand \href@noop [0]{\@secondoftwo}%
\providecommand \href [0]{\begingroup \@sanitize@url \@href}%
\providecommand \@href[1]{\@@startlink{#1}\@@href}%
\providecommand \@@href[1]{\endgroup#1\@@endlink}%
\providecommand \@sanitize@url [0]{\catcode `\\12\catcode `\$12\catcode
  `\&12\catcode `\#12\catcode `\^12\catcode `\_12\catcode `\%12\relax}%
\providecommand \@@startlink[1]{}%
\providecommand \@@endlink[0]{}%
\providecommand \url  [0]{\begingroup\@sanitize@url \@url }%
\providecommand \@url [1]{\endgroup\@href {#1}{\urlprefix }}%
\providecommand \urlprefix  [0]{URL }%
\providecommand \Eprint [0]{\href }%
\providecommand \doibase [0]{https://doi.org/}%
\providecommand \selectlanguage [0]{\@gobble}%
\providecommand \bibinfo  [0]{\@secondoftwo}%
\providecommand \bibfield  [0]{\@secondoftwo}%
\providecommand \translation [1]{[#1]}%
\providecommand \BibitemOpen [0]{}%
\providecommand \bibitemStop [0]{}%
\providecommand \bibitemNoStop [0]{.\EOS\space}%
\providecommand \EOS [0]{\spacefactor3000\relax}%
\providecommand \BibitemShut  [1]{\csname bibitem#1\endcsname}%
\let\auto@bib@innerbib\@empty
%</preamble>
\bibitem [{\citenamefont {Nakamura}\ \emph {et~al.}(2002)\citenamefont
  {Nakamura}, \citenamefont {Pashkin}, \citenamefont {Yamamoto},\ and\
  \citenamefont {Tsai}}]{Nakamura2002}%
  \BibitemOpen
  \bibfield  {author} {\bibinfo {author} {\bibfnamefont {Y.}~\bibnamefont
  {Nakamura}}, \bibinfo {author} {\bibfnamefont {Y.~A.}\ \bibnamefont
  {Pashkin}}, \bibinfo {author} {\bibfnamefont {T.}~\bibnamefont {Yamamoto}},\
  and\ \bibinfo {author} {\bibfnamefont {J.~S.}\ \bibnamefont {Tsai}},\
  }\bibfield  {title} {\bibinfo {title} {{Charge Echo in a Cooper-Pair Box}},\
  }\href {https://doi.org/10.1103/PhysRevLett.88.047901} {\bibfield  {journal}
  {\bibinfo  {journal} {Phys. Rev. Lett.}\ }\textbf {\bibinfo {volume} {88}},\
  \bibinfo {pages} {047901} (\bibinfo {year} {2002})}\BibitemShut {NoStop}%
\bibitem [{\citenamefont {Astafiev}\ \emph {et~al.}(2004)\citenamefont
  {Astafiev}, \citenamefont {Pashkin}, \citenamefont {Nakamura}, \citenamefont
  {Yamamoto},\ and\ \citenamefont {Tsai}}]{Astafiev2004}%
  \BibitemOpen
  \bibfield  {author} {\bibinfo {author} {\bibfnamefont {O.}~\bibnamefont
  {Astafiev}}, \bibinfo {author} {\bibfnamefont {Y.~A.}\ \bibnamefont
  {Pashkin}}, \bibinfo {author} {\bibfnamefont {Y.}~\bibnamefont {Nakamura}},
  \bibinfo {author} {\bibfnamefont {T.}~\bibnamefont {Yamamoto}},\ and\
  \bibinfo {author} {\bibfnamefont {J.~S.}\ \bibnamefont {Tsai}},\ }\bibfield
  {title} {\bibinfo {title} {Quantum noise in the josephson charge qubit},\
  }\href {https://doi.org/10.1103/PhysRevLett.93.267007} {\bibfield  {journal}
  {\bibinfo  {journal} {Phys. Rev. Lett.}\ }\textbf {\bibinfo {volume} {93}},\
  \bibinfo {pages} {267007} (\bibinfo {year} {2004})}\BibitemShut {NoStop}%
\bibitem [{\citenamefont {Yoshihara}\ \emph {et~al.}(2006)\citenamefont
  {Yoshihara}, \citenamefont {Harrabi}, \citenamefont {Niskanen}, \citenamefont
  {Nakamura},\ and\ \citenamefont {Tsai}}]{Yoshihara2006}%
  \BibitemOpen
  \bibfield  {author} {\bibinfo {author} {\bibfnamefont {F.}~\bibnamefont
  {Yoshihara}}, \bibinfo {author} {\bibfnamefont {K.}~\bibnamefont {Harrabi}},
  \bibinfo {author} {\bibfnamefont {A.~O.}\ \bibnamefont {Niskanen}}, \bibinfo
  {author} {\bibfnamefont {Y.}~\bibnamefont {Nakamura}},\ and\ \bibinfo
  {author} {\bibfnamefont {J.~S.}\ \bibnamefont {Tsai}},\ }\bibfield  {title}
  {\bibinfo {title} {{Decoherence of Flux Qubits due to $1/f$ Flux Noise}},\
  }\href {https://doi.org/10.1103/PhysRevLett.97.167001} {\bibfield  {journal}
  {\bibinfo  {journal} {Phys. Rev. Lett.}\ }\textbf {\bibinfo {volume} {97}},\
  \bibinfo {pages} {167001} (\bibinfo {year} {2006})}\BibitemShut {NoStop}%
\bibitem [{\citenamefont {Kumar}\ \emph {et~al.}(2016)\citenamefont {Kumar},
  \citenamefont {Sendelbach}, \citenamefont {Beck}, \citenamefont {Freeland},
  \citenamefont {Wang}, \citenamefont {Wang}, \citenamefont {Yu}, \citenamefont
  {Wu}, \citenamefont {Pappas},\ and\ \citenamefont {McDermott}}]{Kumar2016b}%
  \BibitemOpen
  \bibfield  {author} {\bibinfo {author} {\bibfnamefont {P.}~\bibnamefont
  {Kumar}}, \bibinfo {author} {\bibfnamefont {S.}~\bibnamefont {Sendelbach}},
  \bibinfo {author} {\bibfnamefont {M.~A.}\ \bibnamefont {Beck}}, \bibinfo
  {author} {\bibfnamefont {J.~W.}\ \bibnamefont {Freeland}}, \bibinfo {author}
  {\bibfnamefont {Z.}~\bibnamefont {Wang}}, \bibinfo {author} {\bibfnamefont
  {H.}~\bibnamefont {Wang}}, \bibinfo {author} {\bibfnamefont {C.~C.}\
  \bibnamefont {Yu}}, \bibinfo {author} {\bibfnamefont {R.~Q.}\ \bibnamefont
  {Wu}}, \bibinfo {author} {\bibfnamefont {D.~P.}\ \bibnamefont {Pappas}},\
  and\ \bibinfo {author} {\bibfnamefont {R.}~\bibnamefont {McDermott}},\
  }\bibfield  {title} {\bibinfo {title} {{Origin and Reduction of $1/f$
  Magnetic Flux Noise in Superconducting Devices}},\ }\href
  {https://doi.org/10.1103/PhysRevApplied.6.041001} {\bibfield  {journal}
  {\bibinfo  {journal} {Phys. Rev. Applied}\ }\textbf {\bibinfo {volume} {6}},\
  \bibinfo {pages} {041001} (\bibinfo {year} {2016})}\BibitemShut {NoStop}%
\bibitem [{\citenamefont {{Van Harlingen}}\ \emph {et~al.}(2004)\citenamefont
  {{Van Harlingen}}, \citenamefont {Robertson}, \citenamefont {Plourde},
  \citenamefont {Reichardt}, \citenamefont {Crane},\ and\ \citenamefont
  {Clarke}}]{VanHarlingen2004}%
  \BibitemOpen
  \bibfield  {author} {\bibinfo {author} {\bibfnamefont {D.~J.}\ \bibnamefont
  {{Van Harlingen}}}, \bibinfo {author} {\bibfnamefont {T.~L.}\ \bibnamefont
  {Robertson}}, \bibinfo {author} {\bibfnamefont {B.~L.~T.}\ \bibnamefont
  {Plourde}}, \bibinfo {author} {\bibfnamefont {P.~A.}\ \bibnamefont
  {Reichardt}}, \bibinfo {author} {\bibfnamefont {T.~A.}\ \bibnamefont
  {Crane}},\ and\ \bibinfo {author} {\bibfnamefont {J.}~\bibnamefont
  {Clarke}},\ }\bibfield  {title} {\bibinfo {title} {{Decoherence in
  Josephson-junction qubits due to critical-current fluctuations}},\ }\href
  {https://doi.org/10.1103/PhysRevB.70.064517} {\bibfield  {journal} {\bibinfo
  {journal} {Phys. Rev. B}\ }\textbf {\bibinfo {volume} {70}},\ \bibinfo
  {pages} {064517} (\bibinfo {year} {2004})}\BibitemShut {NoStop}%
\bibitem [{\citenamefont {Martinis}\ \emph {et~al.}(2009)\citenamefont
  {Martinis}, \citenamefont {Ansmann},\ and\ \citenamefont
  {Aumentado}}]{Martinis2009}%
  \BibitemOpen
  \bibfield  {author} {\bibinfo {author} {\bibfnamefont {J.~M.}\ \bibnamefont
  {Martinis}}, \bibinfo {author} {\bibfnamefont {M.}~\bibnamefont {Ansmann}},\
  and\ \bibinfo {author} {\bibfnamefont {J.}~\bibnamefont {Aumentado}},\
  }\bibfield  {title} {\bibinfo {title} {{Energy Decay in Superconducting
  Josephson-Junction Qubits from Nonequilibrium Quasiparticle Excitations}},\
  }\href {https://doi.org/10.1103/PhysRevLett.103.097002} {\bibfield  {journal}
  {\bibinfo  {journal} {Phys. Rev. Lett.}\ }\textbf {\bibinfo {volume} {103}},\
  \bibinfo {pages} {097002} (\bibinfo {year} {2009})}\BibitemShut {NoStop}%
\bibitem [{\citenamefont {Koch}\ \emph {et~al.}(2007)\citenamefont {Koch},
  \citenamefont {Yu}, \citenamefont {Gambetta}, \citenamefont {Houck},
  \citenamefont {Schuster}, \citenamefont {Majer}, \citenamefont {Blais},
  \citenamefont {Devoret}, \citenamefont {Girvin},\ and\ \citenamefont
  {Schoelkopf}}]{Koch2007}%
  \BibitemOpen
  \bibfield  {author} {\bibinfo {author} {\bibfnamefont {J.}~\bibnamefont
  {Koch}}, \bibinfo {author} {\bibfnamefont {T.~M.}\ \bibnamefont {Yu}},
  \bibinfo {author} {\bibfnamefont {J.}~\bibnamefont {Gambetta}}, \bibinfo
  {author} {\bibfnamefont {A.~A.}\ \bibnamefont {Houck}}, \bibinfo {author}
  {\bibfnamefont {D.~I.}\ \bibnamefont {Schuster}}, \bibinfo {author}
  {\bibfnamefont {J.}~\bibnamefont {Majer}}, \bibinfo {author} {\bibfnamefont
  {A.}~\bibnamefont {Blais}}, \bibinfo {author} {\bibfnamefont {M.~H.}\
  \bibnamefont {Devoret}}, \bibinfo {author} {\bibfnamefont {S.~M.}\
  \bibnamefont {Girvin}},\ and\ \bibinfo {author} {\bibfnamefont {R.~J.}\
  \bibnamefont {Schoelkopf}},\ }\bibfield  {title} {\bibinfo {title}
  {{Charge-insensitive qubit design derived from the Cooper pair box}},\ }\href
  {https://doi.org/10.1103/PhysRevA.76.042319} {\bibfield  {journal} {\bibinfo
  {journal} {Phys. Rev. A}\ }\textbf {\bibinfo {volume} {76}},\ \bibinfo
  {pages} {042319} (\bibinfo {year} {2007})}\BibitemShut {NoStop}%
\bibitem [{\citenamefont {Manucharyan}\ \emph {et~al.}(2009)\citenamefont
  {Manucharyan}, \citenamefont {Koch}, \citenamefont {Glazman},\ and\
  \citenamefont {Devoret}}]{Manucharyan2009}%
  \BibitemOpen
  \bibfield  {author} {\bibinfo {author} {\bibfnamefont {V.~E.}\ \bibnamefont
  {Manucharyan}}, \bibinfo {author} {\bibfnamefont {J.}~\bibnamefont {Koch}},
  \bibinfo {author} {\bibfnamefont {L.~I.}\ \bibnamefont {Glazman}},\ and\
  \bibinfo {author} {\bibfnamefont {M.~H.}\ \bibnamefont {Devoret}},\
  }\bibfield  {title} {\bibinfo {title} {{Fluxonium: Single Cooper-Pair Circuit
  Free of Charge Offsets}},\ }\href {https://doi.org/10.1126/science.1175552}
  {\bibfield  {journal} {\bibinfo  {journal} {Science}\ }\textbf {\bibinfo
  {volume} {326}},\ \bibinfo {pages} {113} (\bibinfo {year}
  {2009})}\BibitemShut {NoStop}%
\bibitem [{\citenamefont {Earnest}\ \emph {et~al.}(2018)\citenamefont
  {Earnest}, \citenamefont {Chakram}, \citenamefont {Lu}, \citenamefont
  {Irons}, \citenamefont {Naik}, \citenamefont {Leung}, \citenamefont {Ocola},
  \citenamefont {Czaplewski}, \citenamefont {Baker}, \citenamefont {Lawrence},
  \citenamefont {Koch},\ and\ \citenamefont {Schuster}}]{Earnest2018}%
  \BibitemOpen
  \bibfield  {author} {\bibinfo {author} {\bibfnamefont {N.}~\bibnamefont
  {Earnest}}, \bibinfo {author} {\bibfnamefont {S.}~\bibnamefont {Chakram}},
  \bibinfo {author} {\bibfnamefont {Y.}~\bibnamefont {Lu}}, \bibinfo {author}
  {\bibfnamefont {N.}~\bibnamefont {Irons}}, \bibinfo {author} {\bibfnamefont
  {R.~K.}\ \bibnamefont {Naik}}, \bibinfo {author} {\bibfnamefont
  {N.}~\bibnamefont {Leung}}, \bibinfo {author} {\bibfnamefont
  {L.}~\bibnamefont {Ocola}}, \bibinfo {author} {\bibfnamefont {D.~A.}\
  \bibnamefont {Czaplewski}}, \bibinfo {author} {\bibfnamefont
  {B.}~\bibnamefont {Baker}}, \bibinfo {author} {\bibfnamefont
  {J.}~\bibnamefont {Lawrence}}, \bibinfo {author} {\bibfnamefont
  {J.}~\bibnamefont {Koch}},\ and\ \bibinfo {author} {\bibfnamefont {D.~I.}\
  \bibnamefont {Schuster}},\ }\bibfield  {title} {\bibinfo {title}
  {{Realization of a $\mathrm{\ensuremath{\Lambda}}$ System with Metastable
  States of a Capacitively Shunted Fluxonium}},\ }\href
  {https://doi.org/10.1103/PhysRevLett.120.150504} {\bibfield  {journal}
  {\bibinfo  {journal} {Phys. Rev. Lett.}\ }\textbf {\bibinfo {volume} {120}},\
  \bibinfo {pages} {150504} (\bibinfo {year} {2018})}\BibitemShut {NoStop}%
\bibitem [{\citenamefont {Nguyen}\ \emph {et~al.}(2019)\citenamefont {Nguyen},
  \citenamefont {Lin}, \citenamefont {Somoroff}, \citenamefont {Mencia},
  \citenamefont {Grabon},\ and\ \citenamefont {Manucharyan}}]{Nguyen2018}%
  \BibitemOpen
  \bibfield  {author} {\bibinfo {author} {\bibfnamefont {L.~B.}\ \bibnamefont
  {Nguyen}}, \bibinfo {author} {\bibfnamefont {Y.-H.}\ \bibnamefont {Lin}},
  \bibinfo {author} {\bibfnamefont {A.}~\bibnamefont {Somoroff}}, \bibinfo
  {author} {\bibfnamefont {R.}~\bibnamefont {Mencia}}, \bibinfo {author}
  {\bibfnamefont {N.}~\bibnamefont {Grabon}},\ and\ \bibinfo {author}
  {\bibfnamefont {V.~E.}\ \bibnamefont {Manucharyan}},\ }\bibfield  {title}
  {\bibinfo {title} {{High-Coherence Fluxonium Qubit}},\ }\href
  {https://doi.org/10.1103/PhysRevX.9.041041} {\bibfield  {journal} {\bibinfo
  {journal} {Phys. Rev. X}\ }\textbf {\bibinfo {volume} {9}},\ \bibinfo {pages}
  {041041} (\bibinfo {year} {2019})}\BibitemShut {NoStop}%
\bibitem [{\citenamefont {Brooks}\ \emph {et~al.}(2013)\citenamefont {Brooks},
  \citenamefont {Kitaev},\ and\ \citenamefont {Preskill}}]{Brooks2013}%
  \BibitemOpen
  \bibfield  {author} {\bibinfo {author} {\bibfnamefont {P.}~\bibnamefont
  {Brooks}}, \bibinfo {author} {\bibfnamefont {A.}~\bibnamefont {Kitaev}},\
  and\ \bibinfo {author} {\bibfnamefont {J.}~\bibnamefont {Preskill}},\
  }\bibfield  {title} {\bibinfo {title} {{Protected gates for superconducting
  qubits}},\ }\href {https://doi.org/10.1103/PhysRevA.87.052306} {\bibfield
  {journal} {\bibinfo  {journal} {Phys. Rev. A}\ }\textbf {\bibinfo {volume}
  {87}},\ \bibinfo {pages} {052306} (\bibinfo {year} {2013})}\BibitemShut
  {NoStop}%
\bibitem [{\citenamefont {Groszkowski}\ \emph {et~al.}(2018)\citenamefont
  {Groszkowski}, \citenamefont {Paolo}, \citenamefont {Grimsmo}, \citenamefont
  {Blais}, \citenamefont {Schuster}, \citenamefont {Houck},\ and\ \citenamefont
  {Koch}}]{Groszkowski2017a}%
  \BibitemOpen
  \bibfield  {author} {\bibinfo {author} {\bibfnamefont {P.}~\bibnamefont
  {Groszkowski}}, \bibinfo {author} {\bibfnamefont {A.~D.}\ \bibnamefont
  {Paolo}}, \bibinfo {author} {\bibfnamefont {A.~L.}\ \bibnamefont {Grimsmo}},
  \bibinfo {author} {\bibfnamefont {A.}~\bibnamefont {Blais}}, \bibinfo
  {author} {\bibfnamefont {D.~I.}\ \bibnamefont {Schuster}}, \bibinfo {author}
  {\bibfnamefont {A.~A.}\ \bibnamefont {Houck}},\ and\ \bibinfo {author}
  {\bibfnamefont {J.}~\bibnamefont {Koch}},\ }\bibfield  {title} {\bibinfo
  {title} {Coherence properties of the 0-$\uppi$ qubit},\ }\href
  {https://doi.org/10.1088/1367-2630/aab7cd} {\bibfield  {journal} {\bibinfo
  {journal} {New J. Phys.}\ }\textbf {\bibinfo {volume} {20}},\ \bibinfo
  {pages} {043053} (\bibinfo {year} {2018})}\BibitemShut {NoStop}%
\bibitem [{\citenamefont {Paolo}\ \emph {et~al.}(2019)\citenamefont {Paolo},
  \citenamefont {Grimsmo}, \citenamefont {Groszkowski}, \citenamefont {Koch},\
  and\ \citenamefont {Blais}}]{DiPaolo2018a}%
  \BibitemOpen
  \bibfield  {author} {\bibinfo {author} {\bibfnamefont {A.~D.}\ \bibnamefont
  {Paolo}}, \bibinfo {author} {\bibfnamefont {A.~L.}\ \bibnamefont {Grimsmo}},
  \bibinfo {author} {\bibfnamefont {P.}~\bibnamefont {Groszkowski}}, \bibinfo
  {author} {\bibfnamefont {J.}~\bibnamefont {Koch}},\ and\ \bibinfo {author}
  {\bibfnamefont {A.}~\bibnamefont {Blais}},\ }\bibfield  {title} {\bibinfo
  {title} {Control and coherence time enhancement of the 0{\textendash}$\uppi$
  qubit},\ }\href {https://doi.org/10.1088/1367-2630/ab09b0} {\bibfield
  {journal} {\bibinfo  {journal} {New J. Phys.}\ }\textbf {\bibinfo {volume}
  {21}},\ \bibinfo {pages} {043002} (\bibinfo {year} {2019})}\BibitemShut
  {NoStop}%
\bibitem [{\citenamefont {Gyenis}\ \emph {et~al.}(2019)\citenamefont {Gyenis},
  \citenamefont {Mundada}, \citenamefont {Di~Paolo}, \citenamefont {Hazard},
  \citenamefont {You}, \citenamefont {Schuster}, \citenamefont {Koch},
  \citenamefont {Blais},\ and\ \citenamefont {Houck}}]{Gyenis2019}%
  \BibitemOpen
  \bibfield  {author} {\bibinfo {author} {\bibfnamefont {A.}~\bibnamefont
  {Gyenis}}, \bibinfo {author} {\bibfnamefont {P.~S.}\ \bibnamefont {Mundada}},
  \bibinfo {author} {\bibfnamefont {A.}~\bibnamefont {Di~Paolo}}, \bibinfo
  {author} {\bibfnamefont {T.~M.}\ \bibnamefont {Hazard}}, \bibinfo {author}
  {\bibfnamefont {X.}~\bibnamefont {You}}, \bibinfo {author} {\bibfnamefont
  {D.~I.}\ \bibnamefont {Schuster}}, \bibinfo {author} {\bibfnamefont
  {J.}~\bibnamefont {Koch}}, \bibinfo {author} {\bibfnamefont {A.}~\bibnamefont
  {Blais}},\ and\ \bibinfo {author} {\bibfnamefont {A.~A.}\ \bibnamefont
  {Houck}},\ }\bibfield  {title} {\bibinfo {title} {Experimental realization of
  an intrinsically error-protected superconducting qubit},\ }\href
  {http://arxiv.org/abs/1910.07542} {\bibfield  {journal} {\bibinfo  {journal}
  {arXiv:1910.07542}\ } (\bibinfo {year} {2019})}\BibitemShut {NoStop}%
\bibitem [{\citenamefont {Kitaev}(2006)}]{Kitaev2006}%
  \BibitemOpen
  \bibfield  {author} {\bibinfo {author} {\bibfnamefont {A.}~\bibnamefont
  {Kitaev}},\ }\bibfield  {title} {\bibinfo {title} {Protected qubit based on a
  superconducting current mirror},\ }\href
  {http://arxiv.org/abs/cond-mat/0609441} {\bibfield  {journal} {\bibinfo
  {journal} {arXiv:cond-mat/0609441}\ } (\bibinfo {year} {2006})}\BibitemShut
  {NoStop}%
\bibitem [{\citenamefont {Weiss}\ \emph {et~al.}(2019)\citenamefont {Weiss},
  \citenamefont {Li}, \citenamefont {Ferguson},\ and\ \citenamefont
  {Koch}}]{Weiss2019}%
  \BibitemOpen
  \bibfield  {author} {\bibinfo {author} {\bibfnamefont {D.~K.}\ \bibnamefont
  {Weiss}}, \bibinfo {author} {\bibfnamefont {A.~C.~Y.}\ \bibnamefont {Li}},
  \bibinfo {author} {\bibfnamefont {D.~G.}\ \bibnamefont {Ferguson}},\ and\
  \bibinfo {author} {\bibfnamefont {J.}~\bibnamefont {Koch}},\ }\bibfield
  {title} {\bibinfo {title} {{Spectrum and coherence properties of the
  current-mirror qubit}},\ }\href {https://doi.org/10.1103/PhysRevB.100.224507}
  {\bibfield  {journal} {\bibinfo  {journal} {Phys. Rev. B}\ }\textbf {\bibinfo
  {volume} {100}},\ \bibinfo {pages} {224507} (\bibinfo {year}
  {2019})}\BibitemShut {NoStop}%
\bibitem [{\citenamefont {Clerk}\ \emph {et~al.}(2010)\citenamefont {Clerk},
  \citenamefont {Devoret}, \citenamefont {Girvin}, \citenamefont {Marquardt},\
  and\ \citenamefont {Schoelkopf}}]{aash_rmp}%
  \BibitemOpen
  \bibfield  {author} {\bibinfo {author} {\bibfnamefont {A.~A.}\ \bibnamefont
  {Clerk}}, \bibinfo {author} {\bibfnamefont {M.~H.}\ \bibnamefont {Devoret}},
  \bibinfo {author} {\bibfnamefont {S.~M.}\ \bibnamefont {Girvin}}, \bibinfo
  {author} {\bibfnamefont {F.}~\bibnamefont {Marquardt}},\ and\ \bibinfo
  {author} {\bibfnamefont {R.~J.}\ \bibnamefont {Schoelkopf}},\ }\bibfield
  {title} {\bibinfo {title} {Introduction to quantum noise, measurement, and
  amplification},\ }\href {https://doi.org/10.1103/RevModPhys.82.1155}
  {\bibfield  {journal} {\bibinfo  {journal} {Rev. Mod. Phys.}\ }\textbf
  {\bibinfo {volume} {82}},\ \bibinfo {pages} {1155} (\bibinfo {year}
  {2010})}\BibitemShut {NoStop}%
\bibitem [{\citenamefont {M{\"{u}}ller}\ \emph {et~al.}(2019)\citenamefont
  {M{\"{u}}ller}, \citenamefont {Cole},\ and\ \citenamefont
  {Lisenfeld}}]{Clemens2019}%
  \BibitemOpen
  \bibfield  {author} {\bibinfo {author} {\bibfnamefont {C.}~\bibnamefont
  {M{\"{u}}ller}}, \bibinfo {author} {\bibfnamefont {J.~H.}\ \bibnamefont
  {Cole}},\ and\ \bibinfo {author} {\bibfnamefont {J.}~\bibnamefont
  {Lisenfeld}},\ }\bibfield  {title} {\bibinfo {title} {{Towards understanding
  two-level-systems in amorphous solids: insights from quantum circuits}},\
  }\href {https://doi.org/10.1088/1361-6633/ab3a7e} {\bibfield  {journal}
  {\bibinfo  {journal} {Rep. Prog. Phys.}\ }\textbf {\bibinfo {volume} {82}},\
  \bibinfo {pages} {124501} (\bibinfo {year} {2019})}\BibitemShut {NoStop}%
\bibitem [{\citenamefont {Shnirman}\ \emph {et~al.}(2005)\citenamefont
  {Shnirman}, \citenamefont {Sch\"on}, \citenamefont {Martin},\ and\
  \citenamefont {Makhlin}}]{shnirman2005low}%
  \BibitemOpen
  \bibfield  {author} {\bibinfo {author} {\bibfnamefont {A.}~\bibnamefont
  {Shnirman}}, \bibinfo {author} {\bibfnamefont {G.}~\bibnamefont {Sch\"on}},
  \bibinfo {author} {\bibfnamefont {I.}~\bibnamefont {Martin}},\ and\ \bibinfo
  {author} {\bibfnamefont {Y.}~\bibnamefont {Makhlin}},\ }\bibfield  {title}
  {\bibinfo {title} {{Low- and High-Frequency Noise from Coherent Two-Level
  Systems}},\ }\href {https://doi.org/10.1103/PhysRevLett.94.127002} {\bibfield
   {journal} {\bibinfo  {journal} {Phys. Rev. Lett.}\ }\textbf {\bibinfo
  {volume} {94}},\ \bibinfo {pages} {127002} (\bibinfo {year}
  {2005})}\BibitemShut {NoStop}%
\bibitem [{\citenamefont {Schriefl}\ \emph {et~al.}(2006)\citenamefont
  {Schriefl}, \citenamefont {Makhlin}, \citenamefont {Shnirman},\ and\
  \citenamefont {Sch{\"{o}}n}}]{Schriefl2006}%
  \BibitemOpen
  \bibfield  {author} {\bibinfo {author} {\bibfnamefont {J.}~\bibnamefont
  {Schriefl}}, \bibinfo {author} {\bibfnamefont {Y.}~\bibnamefont {Makhlin}},
  \bibinfo {author} {\bibfnamefont {A.}~\bibnamefont {Shnirman}},\ and\
  \bibinfo {author} {\bibfnamefont {G.}~\bibnamefont {Sch{\"{o}}n}},\
  }\bibfield  {title} {\bibinfo {title} {{Decoherence from ensembles of
  two-level fluctuators}},\ }\href {https://doi.org/10.1088/1367-2630/8/1/001}
  {\bibfield  {journal} {\bibinfo  {journal} {New J. Phys.}\ }\textbf {\bibinfo
  {volume} {8}},\ \bibinfo {pages} {1} (\bibinfo {year} {2006})}\BibitemShut
  {NoStop}%
\bibitem [{\citenamefont {Constantin}\ \emph {et~al.}(2009)\citenamefont
  {Constantin}, \citenamefont {Yu},\ and\ \citenamefont
  {Martinis}}]{Constantin2009}%
  \BibitemOpen
  \bibfield  {author} {\bibinfo {author} {\bibfnamefont {M.}~\bibnamefont
  {Constantin}}, \bibinfo {author} {\bibfnamefont {C.~C.}\ \bibnamefont {Yu}},\
  and\ \bibinfo {author} {\bibfnamefont {J.~M.}\ \bibnamefont {Martinis}},\
  }\bibfield  {title} {\bibinfo {title} {{Saturation of two-level systems and
  charge noise in Josephson junction qubits}},\ }\href
  {https://doi.org/10.1103/PhysRevB.79.094520} {\bibfield  {journal} {\bibinfo
  {journal} {Phys. Rev. B}\ }\textbf {\bibinfo {volume} {79}},\ \bibinfo
  {pages} {094520} (\bibinfo {year} {2009})}\BibitemShut {NoStop}%
\bibitem [{\citenamefont {M{\"{u}}ller}\ \emph {et~al.}(2015)\citenamefont
  {M{\"{u}}ller}, \citenamefont {Lisenfeld}, \citenamefont {Shnirman},\ and\
  \citenamefont {Poletto}}]{Muller2015}%
  \BibitemOpen
  \bibfield  {author} {\bibinfo {author} {\bibfnamefont {C.}~\bibnamefont
  {M{\"{u}}ller}}, \bibinfo {author} {\bibfnamefont {J.}~\bibnamefont
  {Lisenfeld}}, \bibinfo {author} {\bibfnamefont {A.}~\bibnamefont
  {Shnirman}},\ and\ \bibinfo {author} {\bibfnamefont {S.}~\bibnamefont
  {Poletto}},\ }\bibfield  {title} {\bibinfo {title} {{Interacting two-level
  defects as sources of fluctuating high-frequency noise in superconducting
  circuits}},\ }\href {https://doi.org/10.1103/PhysRevB.92.035442} {\bibfield
  {journal} {\bibinfo  {journal} {Phys. Rev. B}\ }\textbf {\bibinfo {volume}
  {92}},\ \bibinfo {pages} {035442} (\bibinfo {year} {2015})}\BibitemShut
  {NoStop}%
\bibitem [{\citenamefont {Dutta}\ and\ \citenamefont {Horn}(1981)}]{Dutta1981}%
  \BibitemOpen
  \bibfield  {author} {\bibinfo {author} {\bibfnamefont {P.}~\bibnamefont
  {Dutta}}\ and\ \bibinfo {author} {\bibfnamefont {P.~M.}\ \bibnamefont
  {Horn}},\ }\bibfield  {title} {\bibinfo {title} {Low-frequency fluctuations
  in solids: $1/f$ noise},\ }\href {https://doi.org/10.1103/RevModPhys.53.497}
  {\bibfield  {journal} {\bibinfo  {journal} {Rev. Mod. Phys.}\ }\textbf
  {\bibinfo {volume} {53}},\ \bibinfo {pages} {497} (\bibinfo {year}
  {1981})}\BibitemShut {NoStop}%
\bibitem [{\citenamefont {Weissman}(1988)}]{Weissman1988}%
  \BibitemOpen
  \bibfield  {author} {\bibinfo {author} {\bibfnamefont {M.~B.}\ \bibnamefont
  {Weissman}},\ }\bibfield  {title} {\bibinfo {title} {$1/f$ noise and other
  slow, nonexponential kinetics in condensed matter},\ }\href
  {https://doi.org/10.1103/RevModPhys.60.537} {\bibfield  {journal} {\bibinfo
  {journal} {Rev. Mod. Phys.}\ }\textbf {\bibinfo {volume} {60}},\ \bibinfo
  {pages} {537} (\bibinfo {year} {1988})}\BibitemShut {NoStop}%
\bibitem [{\citenamefont {Martinis}\ \emph {et~al.}(2005)\citenamefont
  {Martinis}, \citenamefont {Cooper}, \citenamefont {McDermott}, \citenamefont
  {Steffen}, \citenamefont {Ansmann}, \citenamefont {Osborn}, \citenamefont
  {Cicak}, \citenamefont {Oh}, \citenamefont {Pappas}, \citenamefont
  {Simmonds},\ and\ \citenamefont {Yu}}]{Martinis2005}%
  \BibitemOpen
  \bibfield  {author} {\bibinfo {author} {\bibfnamefont {J.~M.}\ \bibnamefont
  {Martinis}}, \bibinfo {author} {\bibfnamefont {K.~B.}\ \bibnamefont
  {Cooper}}, \bibinfo {author} {\bibfnamefont {R.}~\bibnamefont {McDermott}},
  \bibinfo {author} {\bibfnamefont {M.}~\bibnamefont {Steffen}}, \bibinfo
  {author} {\bibfnamefont {M.}~\bibnamefont {Ansmann}}, \bibinfo {author}
  {\bibfnamefont {K.~D.}\ \bibnamefont {Osborn}}, \bibinfo {author}
  {\bibfnamefont {K.}~\bibnamefont {Cicak}}, \bibinfo {author} {\bibfnamefont
  {S.}~\bibnamefont {Oh}}, \bibinfo {author} {\bibfnamefont {D.~P.}\
  \bibnamefont {Pappas}}, \bibinfo {author} {\bibfnamefont {R.~W.}\
  \bibnamefont {Simmonds}},\ and\ \bibinfo {author} {\bibfnamefont {C.~C.}\
  \bibnamefont {Yu}},\ }\bibfield  {title} {\bibinfo {title} {{Decoherence in
  Josephson Qubits from Dielectric Loss}},\ }\href
  {https://doi.org/10.1103/PhysRevLett.95.210503} {\bibfield  {journal}
  {\bibinfo  {journal} {Phys. Rev. Lett.}\ }\textbf {\bibinfo {volume} {95}},\
  \bibinfo {pages} {210503} (\bibinfo {year} {2005})}\BibitemShut {NoStop}%
\bibitem [{\citenamefont {Astafiev}\ \emph {et~al.}(2006)\citenamefont
  {Astafiev}, \citenamefont {Pashkin}, \citenamefont {Nakamura}, \citenamefont
  {Yamamoto},\ and\ \citenamefont {Tsai}}]{Astafiev2006}%
  \BibitemOpen
  \bibfield  {author} {\bibinfo {author} {\bibfnamefont {O.}~\bibnamefont
  {Astafiev}}, \bibinfo {author} {\bibfnamefont {Y.~A.}\ \bibnamefont
  {Pashkin}}, \bibinfo {author} {\bibfnamefont {Y.}~\bibnamefont {Nakamura}},
  \bibinfo {author} {\bibfnamefont {T.}~\bibnamefont {Yamamoto}},\ and\
  \bibinfo {author} {\bibfnamefont {J.~S.}\ \bibnamefont {Tsai}},\ }\bibfield
  {title} {\bibinfo {title} {{Temperature Square Dependence of the Low
  Frequency $1/f$ Charge Noise in the Josephson Junction Qubits}},\ }\href
  {https://doi.org/10.1103/PhysRevLett.96.137001} {\bibfield  {journal}
  {\bibinfo  {journal} {Phys. Rev. Lett.}\ }\textbf {\bibinfo {volume} {96}},\
  \bibinfo {pages} {137001} (\bibinfo {year} {2006})}\BibitemShut {NoStop}%
\bibitem [{\citenamefont {Paladino}\ \emph {et~al.}(2014)\citenamefont
  {Paladino}, \citenamefont {Galperin}, \citenamefont {Falci},\ and\
  \citenamefont {Altshuler}}]{Paladino2014}%
  \BibitemOpen
  \bibfield  {author} {\bibinfo {author} {\bibfnamefont {E.}~\bibnamefont
  {Paladino}}, \bibinfo {author} {\bibfnamefont {Y.~M.}\ \bibnamefont
  {Galperin}}, \bibinfo {author} {\bibfnamefont {G.}~\bibnamefont {Falci}},\
  and\ \bibinfo {author} {\bibfnamefont {B.~L.}\ \bibnamefont {Altshuler}},\
  }\bibfield  {title} {\bibinfo {title} {$1/f$ noise: Implications for
  solid-state quantum information},\ }\href
  {https://doi.org/10.1103/RevModPhys.86.361} {\bibfield  {journal} {\bibinfo
  {journal} {Rev. Mod. Phys.}\ }\textbf {\bibinfo {volume} {86}},\ \bibinfo
  {pages} {361} (\bibinfo {year} {2014})}\BibitemShut {NoStop}%
\bibitem [{\citenamefont {Callen}\ and\ \citenamefont
  {Welton}(1951)}]{PhysRev.83.34}%
  \BibitemOpen
  \bibfield  {author} {\bibinfo {author} {\bibfnamefont {H.~B.}\ \bibnamefont
  {Callen}}\ and\ \bibinfo {author} {\bibfnamefont {T.~A.}\ \bibnamefont
  {Welton}},\ }\bibfield  {title} {\bibinfo {title} {Irreversibility and
  generalized noise},\ }\href {https://doi.org/10.1103/PhysRev.83.34}
  {\bibfield  {journal} {\bibinfo  {journal} {Phys. Rev.}\ }\textbf {\bibinfo
  {volume} {83}},\ \bibinfo {pages} {34} (\bibinfo {year} {1951})}\BibitemShut
  {NoStop}%
\bibitem [{\citenamefont {Clerk}\ \emph {et~al.}(2002)\citenamefont {Clerk},
  \citenamefont {Girvin}, \citenamefont {Nguyen},\ and\ \citenamefont
  {Stone}}]{Clerk2002}%
  \BibitemOpen
  \bibfield  {author} {\bibinfo {author} {\bibfnamefont {A.~A.}\ \bibnamefont
  {Clerk}}, \bibinfo {author} {\bibfnamefont {S.~M.}\ \bibnamefont {Girvin}},
  \bibinfo {author} {\bibfnamefont {A.~K.}\ \bibnamefont {Nguyen}},\ and\
  \bibinfo {author} {\bibfnamefont {A.~D.}\ \bibnamefont {Stone}},\ }\bibfield
  {title} {\bibinfo {title} {{Resonant Cooper-Pair Tunneling: Quantum Noise and
  Measurement Characteristics}},\ }\href
  {https://doi.org/10.1103/PhysRevLett.89.176804} {\bibfield  {journal}
  {\bibinfo  {journal} {Phys. Rev. Lett.}\ }\textbf {\bibinfo {volume} {89}},\
  \bibinfo {pages} {176804} (\bibinfo {year} {2002})}\BibitemShut {NoStop}%
\bibitem [{\citenamefont {Schoelkopf}\ \emph {et~al.}(2003)\citenamefont
  {Schoelkopf}, \citenamefont {Clerk}, \citenamefont {Girvin}, \citenamefont
  {Lehnert},\ and\ \citenamefont {Devoret}}]{Schoelkopf2003}%
  \BibitemOpen
  \bibfield  {author} {\bibinfo {author} {\bibfnamefont {R.~J.}\ \bibnamefont
  {Schoelkopf}}, \bibinfo {author} {\bibfnamefont {A.~A.}\ \bibnamefont
  {Clerk}}, \bibinfo {author} {\bibfnamefont {S.~M.}\ \bibnamefont {Girvin}},
  \bibinfo {author} {\bibfnamefont {K.~W.}\ \bibnamefont {Lehnert}},\ and\
  \bibinfo {author} {\bibfnamefont {M.~H.}\ \bibnamefont {Devoret}},\ }\bibinfo
  {title} {Qubits as spectrometers of quantum noise},\ in\ \href
  {https://doi.org/10.1007/978-94-010-0089-5_9} {\emph {\bibinfo {booktitle}
  {Quantum Noise in Mesoscopic Physics}}},\ \bibinfo {editor} {edited by\
  \bibinfo {editor} {\bibfnamefont {Y.~V.}\ \bibnamefont {Nazarov}}}\ (\bibinfo
   {publisher} {Springer Netherlands},\ \bibinfo {address} {Dordrecht},\
  \bibinfo {year} {2003})\ pp.\ \bibinfo {pages} {175--203}\BibitemShut
  {NoStop}%
\bibitem [{\citenamefont {Halperin}(1976)}]{Temperature}%
  \BibitemOpen
  \bibfield  {author} {\bibinfo {author} {\bibfnamefont {B.~I.}\ \bibnamefont
  {Halperin}},\ }\bibfield  {title} {\bibinfo {title} {Can tunneling levels
  explain the anomalous properties of glasses at very low temperature?},\
  }\href {https://doi.org/10.1111/j.1749-6632.1976.tb39704.x} {\bibfield
  {journal} {\bibinfo  {journal} {Ann. N. Y. Acad. Sci.}\ }\textbf {\bibinfo
  {volume} {279}},\ \bibinfo {pages} {173} (\bibinfo {year}
  {1976})}\BibitemShut {NoStop}%
\bibitem [{\citenamefont {Hunklinger}\ and\ \citenamefont
  {Raychaudhuri}(1986)}]{Hunklinger1986}%
  \BibitemOpen
  \bibfield  {author} {\bibinfo {author} {\bibfnamefont {S.}~\bibnamefont
  {Hunklinger}}\ and\ \bibinfo {author} {\bibfnamefont {A.~K.}\ \bibnamefont
  {Raychaudhuri}},\ }\bibfield  {title} {\bibinfo {title} {{Chapter 3: Thermal
  and Elastic Anomalies in Glasses at Low Temperatures}},\ }\href
  {https://doi.org/10.1016/S0079-6417(08)60015-3} {\bibfield  {journal}
  {\bibinfo  {journal} {Prog. Low Temp. Phys.}\ }\textbf {\bibinfo {volume}
  {9}},\ \bibinfo {pages} {265} (\bibinfo {year} {1986})}\BibitemShut {NoStop}%
\bibitem [{\citenamefont {Phillips}\ and\ \citenamefont
  {Anderson}(1981)}]{phillips1981amorphous}%
  \BibitemOpen
  \bibfield  {author} {\bibinfo {author} {\bibfnamefont {W.~A.}\ \bibnamefont
  {Phillips}}\ and\ \bibinfo {author} {\bibfnamefont {A.}~\bibnamefont
  {Anderson}},\ }\href@noop {} {\emph {\bibinfo {title} {Amorphous solids:
  low-temperature properties}}},\ \bibinfo {series} {Topics in Current
  Physics}, Vol.~\bibinfo {volume} {24}\ (\bibinfo  {publisher}
  {Springer-Verlag Berlin Heidelberg},\ \bibinfo {year} {1981})\BibitemShut
  {NoStop}%
\bibitem [{Note1()}]{Note1}%
  \BibitemOpen
  \bibinfo {note} {In general, longitudinal coupling will modify the linewidth
  of $s_{xx}(\omega )$ in Eq.~\protect \textup {\hbox {\mathsurround \z@
  \protect \normalfont (\ignorespaces \ref {eq:spec}\unskip \@@italiccorr )}}.
  The latter is responsible for the high-frequency regime of $S(\omega )$
  calculated in Sec.~\ref {sec:ens}. Assuming that the TLFs are
  underdamped~\cite {shnirman2005low}, we have confirmed numerically that
  inclusion of longitudinal coupling does not lead to qualitative changes in
  the behavior of $S(\omega )$.}\BibitemShut {Stop}%
\bibitem [{\citenamefont {Faoro}\ and\ \citenamefont
  {Ioffe}(2006)}]{Faoro2006}%
  \BibitemOpen
  \bibfield  {author} {\bibinfo {author} {\bibfnamefont {L.}~\bibnamefont
  {Faoro}}\ and\ \bibinfo {author} {\bibfnamefont {L.~B.}\ \bibnamefont
  {Ioffe}},\ }\bibfield  {title} {\bibinfo {title} {{Quantum Two Level Systems
  and Kondo-Like Traps as Possible Sources of Decoherence in Superconducting
  Qubits}},\ }\href {https://doi.org/10.1103/PhysRevLett.96.047001} {\bibfield
  {journal} {\bibinfo  {journal} {Phys. Rev. Lett.}\ }\textbf {\bibinfo
  {volume} {96}},\ \bibinfo {pages} {047001} (\bibinfo {year}
  {2006})}\BibitemShut {NoStop}%
\bibitem [{\citenamefont {Bloch}(1957)}]{bloch1957generalized}%
  \BibitemOpen
  \bibfield  {author} {\bibinfo {author} {\bibfnamefont {F.}~\bibnamefont
  {Bloch}},\ }\bibfield  {title} {\bibinfo {title} {{Generalized Theory of
  Relaxation}},\ }\href {https://doi.org/10.1103/PhysRev.105.1206} {\bibfield
  {journal} {\bibinfo  {journal} {Phys. Rev.}\ }\textbf {\bibinfo {volume}
  {105}},\ \bibinfo {pages} {1206} (\bibinfo {year} {1957})}\BibitemShut
  {NoStop}%
\bibitem [{\citenamefont {Redfield}(1957)}]{redfield1957theory}%
  \BibitemOpen
  \bibfield  {author} {\bibinfo {author} {\bibfnamefont {A.~G.}\ \bibnamefont
  {Redfield}},\ }\bibfield  {title} {\bibinfo {title} {{On the theory of
  relaxation processes}},\ }\href
  {https://ieeexplore.ieee.org/abstract/document/5392713} {\bibfield  {journal}
  {\bibinfo  {journal} {IBM J. Res. Dev.}\ }\textbf {\bibinfo {volume} {1}},\
  \bibinfo {pages} {19} (\bibinfo {year} {1957})}\BibitemShut {NoStop}%
\bibitem [{\citenamefont {Breuer}\ and\ \citenamefont
  {Petruccione}(2002)}]{breuer2002theory}%
  \BibitemOpen
  \bibfield  {author} {\bibinfo {author} {\bibfnamefont {H.-P.}\ \bibnamefont
  {Breuer}}\ and\ \bibinfo {author} {\bibfnamefont {F.}~\bibnamefont
  {Petruccione}},\ }\href@noop {} {\emph {\bibinfo {title} {The theory of open
  quantum systems}}}\ (\bibinfo  {publisher} {Oxford University Press},\
  \bibinfo {year} {2002})\BibitemShut {NoStop}%
\bibitem [{Note2()}]{Note2}%
  \BibitemOpen
  \bibinfo {note} {In principle, other systems such as a harmonic oscillator
  may be used as a probe. However, this would unnecessarily complicate the
  calculation of interest, and require additional considerations, e.g., of
  leakage into neighboring levels and varying matrix elements for different
  levels. Since the probe system here is merely a calculational tool, its
  particular nature is not of intrinsic interest and we thus choose the
  simplest possible system, i.e., a qubit.}\BibitemShut {Stop}%
\bibitem [{Note3()}]{Note3}%
  \BibitemOpen
  \bibinfo {note} {The same approach applied to the problem of a
  single-electron transistor~\cite {Schoelkopf2003} generates a spectral
  density that is identical to the result obtained by a much more intricate
  calculation based on Keldysh diagrammatics~\cite
  {Johansson2002}.}\BibitemShut {Stop}%
\bibitem [{\citenamefont {Brody}(2014)}]{Brody2014}%
  \BibitemOpen
  \bibfield  {author} {\bibinfo {author} {\bibfnamefont {D.~C.}\ \bibnamefont
  {Brody}},\ }\bibfield  {title} {\bibinfo {title} {{Biorthogonal quantum
  mechanics}},\ }\href {https://doi.org/10.1088/1751-8113/47/3/035305}
  {\bibfield  {journal} {\bibinfo  {journal} {J. Phys. A Math. Theor.}\
  }\textbf {\bibinfo {volume} {47}},\ \bibinfo {pages} {035305} (\bibinfo
  {year} {2014})}\BibitemShut {NoStop}%
\bibitem [{\citenamefont {Johansson}\ \emph {et~al.}(2002)\citenamefont
  {Johansson}, \citenamefont {K{\"{a}}ck},\ and\ \citenamefont
  {Wendin}}]{Johansson2002}%
  \BibitemOpen
  \bibfield  {author} {\bibinfo {author} {\bibfnamefont {G.}~\bibnamefont
  {Johansson}}, \bibinfo {author} {\bibfnamefont {A.}~\bibnamefont
  {K{\"{a}}ck}},\ and\ \bibinfo {author} {\bibfnamefont {G.}~\bibnamefont
  {Wendin}},\ }\bibfield  {title} {\bibinfo {title} {{Full Frequency
  Back-Action Spectrum of a Single-Electron Transistor during Qubit Readout}},\
  }\href {https://doi.org/10.1103/PhysRevLett.88.046802} {\bibfield  {journal}
  {\bibinfo  {journal} {Phys. Rev. Lett.}\ }\textbf {\bibinfo {volume} {88}},\
  \bibinfo {pages} {046802} (\bibinfo {year} {2002})}\BibitemShut {NoStop}%
\bibitem [{Note4()}]{Note4}%
  \BibitemOpen
  \bibinfo {note} {The form of the obtained spectral density in Eq.~\protect
  \textup {\hbox {\mathsurround \z@ \protect \normalfont (\ignorespaces \ref
  {eq:szz}\unskip \@@italiccorr )}} is analogous to $s_{zz}(\omega )$ for a
  single-electron transistor~\cite {Johansson2002,Schoelkopf2003}. While the
  bath interacting with the single-electron transistor is fermionic, the
  relevant excitations are electron--hole pairs characterized by a
  Bose--Einstein distribution. It is thus plausible that the two cases lead to
  similar expressions.}\BibitemShut {Stop}%
\bibitem [{Note5()}]{Note5}%
  \BibitemOpen
  \bibinfo {note} {For a cubic bath spectral function, the TLF's pure-dephasing
  rate vanishes, $\gamma (0)=0$.}\BibitemShut {Stop}%
\bibitem [{\citenamefont {Zener}(1948)}]{Zener1948}%
  \BibitemOpen
  \bibfield  {author} {\bibinfo {author} {\bibfnamefont {C.}~\bibnamefont
  {Zener}},\ }\href@noop {} {\emph {\bibinfo {title} {Elasticity and
  Anelasticity of Metals}}}\ (\bibinfo  {publisher} {University of Chicago
  Press},\ \bibinfo {year} {1948})\BibitemShut {NoStop}%
\bibitem [{Note6()}]{Note6}%
  \BibitemOpen
  \bibinfo {note} {Eq.~\protect \textup {\hbox {\mathsurround \z@ \protect
  \normalfont (\ignorespaces \ref {eq:master}\unskip \@@italiccorr )}} is not
  in the Lindblad form, which in principle disables the unravelling of the
  master equation with quantum trajectory theory. However, the notion of
  processes can still be established from the more complicated diagrammatic
  approach, see Ref.~\protect \rev@citealp {Johansson2002} for
  example.}\BibitemShut {Stop}%
\bibitem [{Note7()}]{Note7}%
  \BibitemOpen
  \bibinfo {note} {Two of the eight possible combinations are ruled out by
  energy conservation.}\BibitemShut {Stop}%
\bibitem [{Note8()}]{Note8}%
  \BibitemOpen
  \bibinfo {note} {If the shift $\delta \omega $ of the spectral density’s
  maxima is small, then a first-order expansion around $\omega =\omega
  _\protect \text {t}$ can be used to obtain the approximation as $\delta
  \omega =\protect \frac {\gamma (\omega _\protect \text {t})}{\gamma '(\omega
  _\protect \text {t})}\leavevmode@ifvmode {\setbox \z@ \hbox {\mathsurround
  \z@ $\nulldelimiterspace \z@ \left (\vcenter to\@ne \big@size {}\right
  .$}\box \z@ }\protect \frac {2}{\protect \sqrt {4+\gamma '(\omega _\protect
  \text {t})^2}}-1\leavevmode@ifvmode {\setbox \z@ \hbox {\mathsurround \z@
  $\nulldelimiterspace \z@ \left )\vcenter to\@ne \big@size {}\right .$}\box
  \z@ }$}\BibitemShut {NoStop}%
\bibitem [{\citenamefont {Sete}\ \emph {et~al.}(2014)\citenamefont {Sete},
  \citenamefont {Gambetta},\ and\ \citenamefont {Korotkov}}]{Sete2014}%
  \BibitemOpen
  \bibfield  {author} {\bibinfo {author} {\bibfnamefont {E.~A.}\ \bibnamefont
  {Sete}}, \bibinfo {author} {\bibfnamefont {J.~M.}\ \bibnamefont {Gambetta}},\
  and\ \bibinfo {author} {\bibfnamefont {A.~N.}\ \bibnamefont {Korotkov}},\
  }\bibfield  {title} {\bibinfo {title} {{Purcell effect with microwave drive:
  Suppression of qubit relaxation rate}},\ }\href
  {https://doi.org/10.1103/PhysRevB.89.104516} {\bibfield  {journal} {\bibinfo
  {journal} {Phys. Rev. B}\ }\textbf {\bibinfo {volume} {89}},\ \bibinfo
  {pages} {104516} (\bibinfo {year} {2014})}\BibitemShut {NoStop}%
\bibitem [{\citenamefont {Itano}\ \emph {et~al.}(1990)\citenamefont {Itano},
  \citenamefont {Heinzen}, \citenamefont {Bollinger},\ and\ \citenamefont
  {Wineland}}]{Itano1990}%
  \BibitemOpen
  \bibfield  {author} {\bibinfo {author} {\bibfnamefont {W.~M.}\ \bibnamefont
  {Itano}}, \bibinfo {author} {\bibfnamefont {D.~J.}\ \bibnamefont {Heinzen}},
  \bibinfo {author} {\bibfnamefont {J.~J.}\ \bibnamefont {Bollinger}},\ and\
  \bibinfo {author} {\bibfnamefont {D.~J.}\ \bibnamefont {Wineland}},\
  }\bibfield  {title} {\bibinfo {title} {{Quantum Zeno effect}},\ }\href
  {https://doi.org/10.1103/PhysRevA.41.2295} {\bibfield  {journal} {\bibinfo
  {journal} {Phys. Rev. A}\ }\textbf {\bibinfo {volume} {41}},\ \bibinfo
  {pages} {2295} (\bibinfo {year} {1990})}\BibitemShut {NoStop}%
\bibitem [{\citenamefont {Phillips}(1972)}]{Phillips1972}%
  \BibitemOpen
  \bibfield  {author} {\bibinfo {author} {\bibfnamefont {W.~A.}\ \bibnamefont
  {Phillips}},\ }\bibfield  {title} {\bibinfo {title} {{Tunneling states in
  amorphous solids}},\ }\href {https://doi.org/10.1007/BF00660072} {\bibfield
  {journal} {\bibinfo  {journal} {J. Low Temp. Phys.}\ }\textbf {\bibinfo
  {volume} {7}},\ \bibinfo {pages} {351} (\bibinfo {year} {1972})}\BibitemShut
  {NoStop}%
\bibitem [{\citenamefont {Ithier}\ \emph {et~al.}(2005)\citenamefont {Ithier},
  \citenamefont {Collin}, \citenamefont {Joyez}, \citenamefont {Meeson},
  \citenamefont {Vion}, \citenamefont {Esteve}, \citenamefont {Chiarello},
  \citenamefont {Shnirman}, \citenamefont {Makhlin}, \citenamefont {Schriefl},\
  and\ \citenamefont {Sch{\"{o}}n}}]{Ithier2005}%
  \BibitemOpen
  \bibfield  {author} {\bibinfo {author} {\bibfnamefont {G.}~\bibnamefont
  {Ithier}}, \bibinfo {author} {\bibfnamefont {E.}~\bibnamefont {Collin}},
  \bibinfo {author} {\bibfnamefont {P.}~\bibnamefont {Joyez}}, \bibinfo
  {author} {\bibfnamefont {P.~J.}\ \bibnamefont {Meeson}}, \bibinfo {author}
  {\bibfnamefont {D.}~\bibnamefont {Vion}}, \bibinfo {author} {\bibfnamefont
  {D.}~\bibnamefont {Esteve}}, \bibinfo {author} {\bibfnamefont
  {F.}~\bibnamefont {Chiarello}}, \bibinfo {author} {\bibfnamefont
  {A.}~\bibnamefont {Shnirman}}, \bibinfo {author} {\bibfnamefont
  {Y.}~\bibnamefont {Makhlin}}, \bibinfo {author} {\bibfnamefont
  {J.}~\bibnamefont {Schriefl}},\ and\ \bibinfo {author} {\bibfnamefont
  {G.}~\bibnamefont {Sch{\"{o}}n}},\ }\bibfield  {title} {\bibinfo {title}
  {{Decoherence in a superconducting quantum bit circuit}},\ }\href
  {https://doi.org/10.1103/PhysRevB.72.134519} {\bibfield  {journal} {\bibinfo
  {journal} {Phys. Rev. B}\ }\textbf {\bibinfo {volume} {72}},\ \bibinfo
  {pages} {134519} (\bibinfo {year} {2005})}\BibitemShut {NoStop}%
\bibitem [{\citenamefont {Shnirman}\ \emph {et~al.}(2007)\citenamefont
  {Shnirman}, \citenamefont {Sch{\"o}n}, \citenamefont {Martin},\ and\
  \citenamefont {Makhlin}}]{shnirmanreview}%
  \BibitemOpen
  \bibfield  {author} {\bibinfo {author} {\bibfnamefont {A.}~\bibnamefont
  {Shnirman}}, \bibinfo {author} {\bibfnamefont {G.}~\bibnamefont {Sch{\"o}n}},
  \bibinfo {author} {\bibfnamefont {I.}~\bibnamefont {Martin}},\ and\ \bibinfo
  {author} {\bibfnamefont {Y.}~\bibnamefont {Makhlin}},\ }\bibfield  {title}
  {\bibinfo {title} {$1/f$ noise and two-level systems in josephson qubits},\
  }in\ \href@noop {} {\emph {\bibinfo {booktitle} {Electron Correlation in New
  Materials and Nanosystems}}},\ \bibinfo {editor} {edited by\ \bibinfo
  {editor} {\bibfnamefont {K.}~\bibnamefont {Scharnberg}}\ and\ \bibinfo
  {editor} {\bibfnamefont {S.}~\bibnamefont {Kruchinin}}}\ (\bibinfo
  {publisher} {Springer Netherlands},\ \bibinfo {address} {Dordrecht},\
  \bibinfo {year} {2007})\ pp.\ \bibinfo {pages} {343--356}\BibitemShut
  {NoStop}%
\bibitem [{Note9()}]{Note9}%
  \BibitemOpen
  \bibinfo {note} {\label {f3}For $j\protect \neq 0,1$, similar observation
  shows that $ c_{j\alpha }$ and $d_{j\mu }$ are of order $\kappa ^1$ and
  $\kappa ^0$, respectively.}\BibitemShut {Stop}%
\end{thebibliography}%

\end{document}